\documentclass[12pt]{article}

\usepackage{amssymb}
\usepackage{latexsym}
\usepackage{amsmath}
\usepackage[dvips]{graphicx}
\usepackage{array}
\usepackage{tabularx}
\usepackage{fancybox}
\usepackage{supertabular}

\usepackage{color}

\newcommand{\mev}{\, \mbox{\rm MeV}}
\newcommand{\gev}{\, \mbox{\rm GeV}}

\newcommand{\lsim}{\stackrel{<}{_\sim}}

\newcommand{\chpt}{$\chi$PT }
\newcommand{\rcht}{R$\chi$T }

\newcommand{\bea}{\begin{eqnarray}}
\newcommand{\eea}{\end{eqnarray}}
\newcommand{\beq}{\begin{equation}}
\newcommand{\eeq}{\end{equation}}

\newcommand{\cO}{{\cal O}}

\newcommand{\ket}{\,\rangle}
\newcommand{\bra}{\langle \,}

\newcommand{\es}{\varepsilon^{\phantom{a}}_{\mathrm{S}}}
\newcommand{\ep}{\varepsilon^{\phantom{a}}_{\mathrm{P}}}
\newcommand{\ppa}{\lambda_1^{\mathrm{PP}}}
\newcommand{\ppb}{\lambda_2^{\mathrm{PP}}}
\newcommand{\ppc}{\lambda_3^{\mathrm{PP}}}
\newcommand{\ssa}{\lambda_1^{\mathrm{SS}}}
\newcommand{\ssb}{\lambda_2^{\mathrm{SS}}}
\newcommand{\ssc}{\lambda_3^{\mathrm{SS}}}
\newcommand{\spa}{\lambda_1^{\mathrm{SP}}}
\newcommand{\spb}{\lambda_2^{\mathrm{SP}}}

\newcommand{\lam}{\lambda}
\newcommand{\bs}{b^{\mathrm{S}}}
\newcommand{\bp}{b^{\mathrm{P}}}
\newcommand{\bsp}{b^{\mathrm{SP}}}
\newcommand{\bsm}{{b^{\mathrm{S}\,\mu}}}
\newcommand{\bpm}{{b^{\mathrm{P}\,\mu}}}
\newcommand{\bspm}{{b^{\mathrm{SP}\,\mu}}}
\newcommand{\bsn}{{b^{\mathrm{S}\,\nu}}}
\newcommand{\bpn}{{b^{\mathrm{P}\,\nu}}}
\newcommand{\bspn}{{b^{\mathrm{SP}\,\nu}}}
\newcommand{\as}{{a^{\mathrm{S}}}}
\newcommand{\ap}{{a^{\mathrm{P}}}}
\newcommand{\asp}{{a^{\mathrm{SP}}}}
\newcommand{\ks}{{k^{\mathrm{S}}}}
\newcommand{\kp}{{k^{\mathrm{P}}}}

\def\mapright#1#2{\smash{
     \mathop{-\!\!\!-\!\!\!\rightarrow}\limits^{#1}_{#2}}}

\newif\ifpdf
\ifx\pdfoutput\undefined
\pdffalse 
\else
\pdfoutput=1 
\pdftrue
\fi

\newcommand{\QCD}{{\rm QCD} }

\begin{document}
\begin{titlepage}

\ifpdf
\DeclareGraphicsExtensions{.pdf, .jpg}
\else
\DeclareGraphicsExtensions{.eps, .jpg}
\fi

\vspace*{-1.8cm}
\begin{flushright}
{\small\sf  IFIC/05-25\\FTUV/05-1004\\}
\end{flushright}

\vspace*{2.5cm} 
\begin{center}
{\Large\bf One-loop Renormalization of Resonance Chiral Theory~:}  \\
$   $ \\
{\Large\bf Scalar and Pseudoscalar Resonances$^*$}
\\[20mm]

{\normalsize\bf \sc I. Rosell$^{1}$, P. Ruiz-Femen\'{\i}a$^2$}
and {\normalsize\bf \sc J. Portol\'es$^1$} \\

\vspace{1.4cm} 

${}^{1)}$ {\em Departament de F\'{\i}sica Te\`orica, IFIC, CSIC --- 
Universitat de Val\`encia \\ 
Edifici d'Instituts de Paterna, Apt. Correus 22085, E-46071 
Val\`encia, Spain} \\[10pt]
${}^{2)}$ {\em Theory Division, Max-Planck-Institut f\"ur Physik, \\
F\"ohringer Ring 6, D-80805 M\"unchen, Germany} \\
\end{center}

\vfill
\begin{abstract}
We consider the Resonance Chiral Theory with one multiplet of 
scalar and pseudoscalar resonances, up to bilinear couplings in the
resonance fields, and evaluate its $\beta$-function at one-loop
with the use of the background field method. Thus
we also provide the full set of operators that renormalize the theory at
one loop and render it finite.

\end{abstract}

\vfill
\noindent 
*~Work supported in part by HPRN-CT2002-00311 (EURIDICE).
\end{titlepage}

\section{Introduction}
It is well known that the study of low--energy hadrodynamics is
tampered with by our present inability to implement non-perturbative
Quantum Chromodynamics (QCD) fully in those processes. 
Hence, conspicuously, in order to disentangle New Physics effects
from the Standard Model we need to work out a frame that enforces
QCD in this energy region as close as possible.
\par
In the very low--energy regime
(typically $E \ll M_V$, where $M_V$ is short for the mass of the
lightest vector meson multiplet) Chiral Perturbation Theory ($\chi$PT)
\cite{Weinberg:1978kz,Gasser:1983yg} has become a successful model-independent 
tool that exploits the main features of an Effective Theory approach
to QCD, namely the symmetries of the underlying theory and the
phenomenologically suited assumption of spontaneous breaking of the
chiral symmetry.
$\chi$PT is a non-decoupling Effective Theory ruled by the 
$SU(N_F)_L \otimes SU(N_F)_R$ chiral symmetry of massless \QCD and it is
built in terms of the
pseudo-Goldstone bosons (to be identified with the lightest multiplet
of pseudoscalar mesons) that are generated in the phase transition 
driven by the spontaneous breaking of the symmetry, all the heavier states
being integrated out. This framework is worked out as a perturbative
expansion in the momenta and masses of the pseudoscalar mesons and
it has proven to be a rigorous and fruitful scheme \cite{Ecker:1994gg}.
Although \chpt is non-renormalizable {\em stricto sensu}, it supports
a perturbative renormalization procedure where all loop divergences
can be reabsorbed (when regularized within a scheme that respects
chiral symmetry) into a finite number of new operators. This procedure
is well defined order by order in the chiral expansion. Accordingly
its generating functional has been systematically calculated up to
${\cal O}(p^6)$, that is two loops for
even--intrinsic--parity processes 
\cite{Gasser:1983yg,Fearing:1994ga,Unterdorfer:2002zg}
and one loop in the odd--intrinsic--parity sector \cite{Akhoury:1990px}.
$\chi$PT has profusely guided light meson dynamics in the last twenty
years.
\par
At higher energies ($M_V \lsim E \lsim 2 \gev$) the situation is more
involved. This regime is populated by many resonances and the absence
of a mass gap in the spectrum of states makes difficult to provide
a formal Effective Theory approach to implement \QCD properly. Hence we 
have to rely on additional information provided by the strong interaction
underlying theory. Large--$N_C$ \QCD \cite{'tHooft:1973jz}
furnishes a practical scenario to work with. The
$N_C \rightarrow \infty$ limit strongly constrains meson dynamics by
asserting that the Green Functions of the theory are described by the
tree diagrams of an effective local Lagrangian with local vertices and
meson fields, higher corrections in $1/N_C$ being yielded by loops
described within the same Lagrangian theory. In addition the spectrum
of this limit of \QCD is given by an infinite set of non-decaying meson
states. This scheme has supplied thorough insights to extract information
from the theory \cite{Manohar:1998xv,Pich:2002xy}. Nevertheless
a strict formulation
of large--$N_C$ \QCD in the $N_C \rightarrow \infty$ limit is still 
lacking, mainly due to the fact that we do not know how to implement
an infinite spectrum in a model--independent manner.
\par
The known phenomenology of hadron processes seems to support the 
assumption that when a resonance leads the dynamics of an observable,
heavier resonances, with the same quantum numbers, tend to play a decreasing
role. This reasonable endorsed conjecture helps us to modelize 
large--$N_C$ \QCD with a Lagrangian theory that involves, besides pseudoscalar
mesons, one $U(3)$ multiplet of scalar, pseudoscalar, vector and 
axial-vector resonances, each one to be identified with the 
lightest corresponding states
in the phenomenological spectrum. The remaining restrictions on the 
theory come from the symmetry properties of QCD, like chiral symmetry
driving the interactions of the pseudoscalar mesons and $U(3)$ unitary
symmetry for the matter fields (resonances) \cite{Coleman:1969sm}.
This
construction, that constitutes the basis of the Resonance Chiral Theory
(R$\chi$T),
was carried out in detail in Ref.~\cite{Ecker:1988te}, though
considering operators with only one resonance field in the
even-intrinsic-parity
sector\footnote{An extension to operators contributing, upon integration,
to the ${\cal O}(p^6)$ \chpt  Lagrangian is under way \cite{nosfuture}.}. 
Notice that
in contrast to many modelizations of the resonance fields that have been
widely employed in the literature, \rcht only uses basic \QCD symmetry 
features without any additional {\em ad hoc} assumptions. Its model
aspect only comes from the fact that we do not include an infinite spectrum
in the theory, which is one of the features of 
the $N_C \rightarrow \infty$ limit of QCD. Finally
the elements that provide the theory, symmetries and spectrum, give the
structure of the operators but do not supply information on the couplings
of the Lagrangian. However these are strongly constrained by 
short-distance properties of the underlying \QCD dynamics~: an interplay
between hadronic form factors and 
Green functions of the \QCD currents can furnish all--important information
on the couplings \cite{Ecker:1989yg} through a matching procedure that
has been very much used lately with notable success 
\cite{Moussallam:1997xx,Cirigliano:2004ue}.
Other modelizations have also been used to fulfill this task 
\cite{Prades:1993ys}.
\par
Since its inception \rcht has been applied both to the study of resonance
contributions in weak interaction processes (radiative and non--leptonic
kaon decays) \cite{Ecker:1992de} and to the study of form factors of mesons
\cite{Guerrero:1997ku} where only the \rcht Lagrangian at tree level has been
used and, accordingly, the leading contribution in the
large--$N_C$ model we are describing has been obtained. The next-to-leading
order in the $1/N_C$ expansion arises from one loop calculations within
the theory and its control starts to be necessary both on grounds of 
the convergence of the predictions and to straighten our knowledge of
non-perturbative QCD. Some pioneering work at one-loop has already been
performed \cite{GomezDumm:2000fz,Rosell:2004mn} showing not only the technical 
difficulties that appear but also the conceptual intricacies that involve
the construction of the theory.
\par
\rcht is non-renormalizable. Moreover the lack of an expansion parameter
in the Lagrangian does not make feasible the application of a perturbative
renormalization program based on a well defined power-counting scheme
analogous to the one in $\chi$PT. Nevertheless
from a practical point of view the situation is similar to the \chpt case
\cite{Buchler:2003vw}.
As shown in Ref.~\cite{Rosell:2004mn}, where the vector form factor of the
pion was calculated at one loop in R$\chi$T, it is possible to construct a 
finite number of operators, within the theory, whose couplings can 
absorb the divergences coming from one loop
diagrams. The only requirement is, of course, that the regularization procedure
of the loop divergences respects the symmetries of the Lagrangian.
\par
In the present article we have studied the full one-loop generating functional
that arises from \rcht when one multiplet of scalar and pseudoscalar 
resonances are considered and only up to 
bilinear couplings in the resonances are
included. We have evaluated the divergent contributions
and, consequently, we have obtained the full set of operators needed to
renormalize the theory properly. The conceptual differences with the
\chpt renormalization program will also be stressed.
\par
In Section~2 we describe shortly the content of \rcht that is of interest
in our case. Section~3 is devoted to explain the procedure and hints that
we follow to perform the evaluation of the generating functional, whose
results are given in Section~4 and commented in Section~5. In Section~6
we point out the conclusions and summarize. Most of the
technical details
are relegated to the Appendices.

\section{\rcht with scalar and pseudoscalar resonances}

We consider the R$\chi$T Lagrangian constituted by pseudo-Goldstone
bosons (the lightest pseudoscalar mesons) and one multiplet of both 
scalar and pseudoscalar resonances. Motivated by the large-$N_C$ limit
we include $U(3)$ multiplets for the spectrum though we limit ourselves
to $SU(3)$ external currents as we are not interested in anomaly related
issues. Our Lagrangian reads~:
\begin{eqnarray}\label{eq:lagr1}
\mathcal{L}_{\mathrm{R}\chi\mathrm{T}}(\phi, \mathrm{S},\mathrm{P})
&=&\mathcal{L}_{\chi}^{(2)}\,+\,
\mathcal{L}_{\mathrm{kin}}(\mathrm{S},\mathrm{P}) \,+\,
\mathcal{L}_{2}(\mathrm{S})\,+\,  \mathcal{L}_{2}(\mathrm{P})\,+
\, \mathcal{L}_{2}(\mathrm{S},\mathrm{P}) \, ,
\end{eqnarray}
where $\mathcal{L}_{\chi}^{(2)}$ is the $\cO(p^2)$ $\chi$PT Lagrangian,
\begin{eqnarray}
\mathcal{L}_{\chi}^{(2)}&=&
\frac{F^2}{4} \bra u_\mu u^\mu \, + \, \chi_+ \ket \,,
\end{eqnarray}
with $F \simeq 92.4 \mev$ the decay constant of the pion in the chiral limit
and the brackets $\langle ... \rangle$ stand for a trace in the flavour
space. The kinetic term for the scalar $S(0^{++})$ and pseudoscalar
$P(0^{-+})$ resonances is~:
\begin{eqnarray} \label{eq:kineticsp}
\mathcal{L}_{\mathrm{kin}} (\mathrm{S},\mathrm{P}) &=& \frac{1}{2} 
\, \sum_{R \, = \, S,P} \bra \nabla^\mu R \,\nabla_\mu R \,-\, M_R^2\, R^2 
\ket  \,, 
\end{eqnarray}
that also includes interaction terms through the covariant derivatives.
Finally the pure interacting terms for resonances and pseudoscalars are
given by \cite{Ecker:1988te,nosfuture}~:
\begin{eqnarray} \label{eq:spint}
\mathcal{L}_{2}(\mathrm{S})&=& c_d \bra S \, u_\mu u^\mu \ket  +
 c_m \bra S \,\chi_+ \ket  
+ \ssa \bra SS \,u^\mu u_\mu \ket + \ssb \bra S u_\mu S u^\mu
 \ket  + \ssc \bra SS\, \chi_+ \ket \,,\,
 \phantom{\frac{1}{2}} \nonumber \\
\mathcal{L}_{2}(\mathrm{P})&=&i\,d_m \bra P \,\chi_- \ket  \,
+\, \ppa \bra PP \,u^\mu u_\mu \ket \,+\, \ppb \bra P u_\mu P u^\mu 
\ket \, +\, \ppc \bra PP\, \chi_+ \ket \,,\,\phantom{\frac{1}{2}} \\
\mathcal{L}_{2}(\mathrm{S},\mathrm{P})&=&\spa\bra \{ \nabla_\mu S, P \} 
u^\mu \ket + i \, \spb \bra \{ S, P \} \chi_- \ket \, , 
\phantom{\frac{1}{2}} \nonumber
\end{eqnarray}
where all the couplings are real valued.
We follow closely the notation of Refs.~\cite{Ecker:1988te,nosfuture} that, 
for convenience, is recalled in Appendix~\ref{app:uno}.
\par
Several comments on our Lagrangian theory are suitable here~:
\begin{enumerate}
\item[-] Notice that we are not including the \chpt Lagrangian of 
$\cO (p^4)$ and higher orders. It has been shown \cite{Ecker:1988te} 
that ${\cal L}_{\chi}^{(4)}$
is largely saturated\footnote{This is much more clear 
in the case of vector and axial-vector
resonances as their phenomenology is better known.} by the resonance 
exchange generated by the linear terms in the resonance field given by
${\cal L}_2(S)$ and ${\cal L}_2(P)$, hence the
explicit introduction of ${\cal L}_{\chi}^{(4)}$ would amount to include an
overlap between both resonance contributions.
An analogous analysis at $\cO(p^6)$ has not been
performed systematically but it also looks a reasonable assumption. 
Thus our theory stands for a complete resonance saturation of the 
\chpt Lagrangian.
\item[-] The structure of the resonance interacting operators follows a
definite pattern. The linear terms in the resonance fields were already
introduced in Ref.~\cite{Ecker:1988te} and have the structure 
$\langle R \, \chi^{(2)} \rangle$, where
$\chi^{(2)} \, \equiv \, \chi_{\pm}, \, u_{\mu}u^{\mu} \,$
is a chiral tensor, involving Goldstone bosons and external currents,
of $\cO(p^2)$ in the chiral counting\footnote{This is in fact the origin
of the subindex 2 in the Lagrangian (${\cal L}_2$),
i.e. that the Goldstone and external
currents are introduced with chiral tensors of that order.}. However
the theory does not look
consistent with these pieces only. This is because the kinetic term
in Eq.~(\ref{eq:kineticsp}), due to the covariant derivative,
also includes an interacting term with two resonances. Hence it seems
congruous to include all the interacting terms with two resonances
that have the structure of the kinetic term, i.e. 
$\langle R \, R \, \chi^{(2)} \rangle$, as we have done \cite{nosfuture}. 

\item[-] The \rcht Lagrangian satisfies, by construction, the strictures
of chiral dynamics at very low-energies ($E \ll M_R$). Notwithstanding, it 
is clear that there is no small coupling or kinematical parameter that could
allow us to perform a perturbative expansion in order to solve the 
effective action of the theory, as it happens in \chpt. Hence it looks that
the exclusion of many (infinite) operators (e.g. one resonance and a 
${\cal O}(p^4)$ chiral tensor, etc), satisfying the required symmetries,
is not justified. The way out to this assessments has several components.
On one side \rcht is not an Effective 
Theory but a consistent phenomenological Lagrangian model where many of
the constraints from \QCD are enforced. In particular it can be seen
\cite{nosfuture} that short-distance conditions limit strongly the operators
that are allowed (those with higher order chiral tensors tend to violate the
\QCD ruled asymptotic behaviour of Green's functions or form factors).
On the other side we have the large-$N_C$
limit in order to guide a loop perturbative expansion, not in the Lagrangian,
but in the observables evaluated with it. 
\par 
It has also been proposed 
\cite{Bruns:2004tj} that, due to the fact that the chiral counting is
spoiled when resonances are included in loops, it could be possible to
keep the chiral counting by disentangling the \lq \lq {\em hard}" modes
that could be absorbed in the
renormalization program. In this way one gets a chiral expansion even if
resonance contributions in the loop are considered. This procedure can
be useful but only if one is interested in the application at very low
energies out of the resonance region. 

\item[-] The Lagrangian theory described by Eq.~(\ref{eq:lagr1}) satisfies
the $N_C$ counting rules \cite{'tHooft:1973jz,Kaiser:2000gs}.
Leading operators in the large-$N_C$ limit have one trace in the flavour
space and we attach to this leading order terms. From the interaction vertices
we see that the couplings $F, c_d, c_m$ and $d_m$ are
of $\cO(\sqrt{N_C})$ while $\lambda_i^{SS}, \lambda_i^{PP}$ and 
$\lambda_i^{SP}$ are of $\cO(1)$. Short-distance constraints on the
asymptotic behaviour of form factors and Green functions 
\cite{Pich:2002xy,Ecker:1989yg} provide, in the $N_C \rightarrow \infty$
limit, the following relations~:
\begin{eqnarray}  \label{eq:largecd} 
c_m \, = \, c_d \, = \, \sqrt{2} \; d_m & = & \frac{F}{2} \; ,  
\end{eqnarray}
and $M_P^2  =  2 \, (1 \, - \, \delta) \, M_S^2 $ with
 $\delta \, \simeq \, 3 \, \pi  \alpha_S \, F^2/M_S^2 \sim \, 0.08 \, \alpha_S$.
Short-distance constraints on the $\lambda_i^{RR}$ couplings in the 
$N_C \rightarrow \infty$ are, at the
moment, more controversial \cite{nosfuture,ffheretic}~:
\begin{eqnarray} \label{eq:largecd2}
\lambda_3^{SS} \, = \, \lambda_3^{PP} & = & 0 \; , \nonumber \\
\lambda_1^{SP} \, = \, 4 \, \lambda_2^{SP} \, = \, - \, \frac{d_m}{c_m} 
& = & - \, \frac{1}{\sqrt{2}} \; , 
\end{eqnarray}
where we have used Eq.~(\ref{eq:largecd}). 
In Appendix~\ref{app:dos} we explain how these last results are obtained. 
Though the relations shown in Eqs.~(\ref{eq:largecd},\ref{eq:largecd2}) could
be used to simplify the outcome
of the calculations presented in this article, we will give the 
full results without 
short-distance constraints built-in so as not to lose generality.
\end{enumerate}
The lack of an expansion coupling or parameter in \rcht hinders a 
perturbative renormalization like the one applied in \chpt. Nevertheless 
a complete one-loop calculation of the vector form factor of the pion
in \rcht \cite{Ecker:1988te} was performed in Ref.~\cite{Rosell:2004mn}
with special attention to the renormalization program. It was shown that,
using dimensional regularization,
all the divergences could be absorbed by the introduction of local
operators fulfilling the symmetry requirements. This is a particular case 
of the well known fact that all divergences are local in a quantum field
theory \cite{Collins}, and are given by a polynomial in the external 
momenta or masses. Hence it is reasonable to consider the construction of the 
full set of operators that renders our ${\cal L}_{R\chi T}(\phi,S,P)$ theory
finite up to one-loop. Accordingly we perform the one-loop
generating functional of our Lagrangian theory to evaluate the full set of
divergences that arise. This we pursue in the rest of the article.

\section{Generating functional at one loop}

The generating functional of the connected Green functions, $W[J]$,
is the logarithm of the vacuum-to-vacuum transition amplitude in the
presence of external sources $J(x)$ coupled to bilinear quark currents~:
\begin{equation}
e^{\,i \, W[J]} \, = \, \frac{1}{{\cal N}} \, \int  \, [\, d \psi \, ] \, \, \, 
e^{\, i \, S_0[\psi,J]} \; ,
\end{equation}
where the normalization is such that $W[0]=0$ and the field $\psi$ is,
in our case, short for the Goldstone and resonance mesons.
The evaluation of the generating functional of our Lagrangian theory
${\cal L}_{R\chi T}(\phi,S,P)$,
is readily done with the background field method \cite{bfm,bfm2},
where the action is expanded around the classical fields $\psi_{cl}$.
By defining the quantum field as 
$\Delta \psi = \psi - \psi_{cl}$, the expansion up to one loop ($L=1$)
is given by~:
\begin{align}
W[J]_{L=1} \, = \, S_0[\psi_{cl},J] \, - \, i \, \ln \left[ \,  
 \int \, [\, d \Delta \psi \, ] \, \exp \left( \phantom{\frac{1}{2}}
\right. \right. & 
\! \! \! \!   i
\int \, d^4x_1 \, \frac{\delta \, S_0[\psi,J]}{\delta \psi_i(x_1)} \, 
\Big|_{\psi_{cl}}
\, \Delta \psi_i(x_1) \, \nonumber \\ 
\! \! \! \! \! \!\! \! \! &  \! \!\! \! \! \! \!\!
\! \! \! \!\! \! \! \! \! \! \!\! \! \! 
\!\! \! \! 
\! \! \! \!\! \! \! \! \! \! \!\! \! \! \!\! \! \! \!\! \! \! \!\! \! \! 
\left. \left.+ \, \frac{i}{2!} \, \int \,
 d^4x_1 \, d^4x_2 \, \, \Delta \psi_i(x_1) \, 
\frac{\delta^2 \, S_0[\psi,J]}{\delta \psi_i(x_1) \, \delta \psi_j(x_2)} \,
 \Big|_{\psi_{cl}} \, \Delta \psi_j(x_2) \, \right) \, 
 \right]   ,
\end{align}
but for an irrelevant constant. The $i,j$ indices run over all the
different fields and are summed over. The classical field $\psi_{cl}$
is, by definition, the solution of~:
\begin{equation} \label{eq:eom}
\frac{\delta S_0[\psi,J]}{\delta \psi_i(x)} \, \Bigg|_{\psi_{cl}} \, = \, 0 \, ,
\end{equation}
that provides the implicit relation $\psi_{cl} = \psi_{cl}[J]$ and
the Equations of Motion (EOM) for the classical fields. 
The explicit expressions of the latter are detailed in 
Appendix~\ref{app:tres}. Solving the
remaining gaussian integral in the Euclidean spacetime and coming back
to Minkowsky we have finally~:
\begin{eqnarray} \label{eq:sl1}
W[J]_{L=1}  &=& S_0[\psi_{cl},J] \, + \, S_1[\psi_{cl},J]\, ,  \\
S_1[\psi_{cl},J] &=& \frac{i}{2} \,
\ln  \,\mbox{det} \, {\cal D}(\psi_{cl},J) 
 \, \;, 
\end{eqnarray}
where $ {\cal D}(\psi_{cl},J)$ is the quadratic differential operator
specified by~:
\begin{equation}
\langle \, x \,  | \, {\cal D}(\psi_{cl},J) \, | \,  y \, 
\rangle_{ij} \, = \,
 \frac{\delta^2 \, S_0[\psi,J]}{\delta \psi_i(x) \, \delta \psi_j(y)} 
\Bigg|_{\psi_{cl}} \; .
\end{equation}
The action at one loop needs regularization and, following the use
within $\chi$PT, we will proceed by working
in $D$ spacetime dimensions, a procedure that preserves the relevant symmetries
of our theory. 
Divergences in the functional integration are local and, within dimensional 
regularization, can be absorbed through local operators that satisfy 
the same symmetries
than the original theory \cite{Collins}. The one-loop renormalized 
Lagrangian is thus defined by~:
\begin{equation} \label{eq:renoo}
 {\cal L}_{1}[\psi,J] \, = \,  
\mu^{D-4} \, \left( \, {\cal L}_{1}^{\mathrm{ren}}[\psi,J; \mu] \, + \, 
 \frac{1}{(4\pi)^2} \, \frac{1}{D-4} \, 
 {\cal L}_{1}^{\mathrm{div}}[\psi,J] \, \right) \, .
\end{equation}
In Eq.~(\ref{eq:renoo}) we
have split the one-loop bare Lagrangian into a renormalized and
 a divergent part, and the scale $\mu$ is
introduced in order to restore the correct dimensions in the
 renormalized Lagrangian for $D\ne 4$. The divergent
part ${\cal L}_{1}^{\mathrm{div}}$ contains the counterterms
 which exactly cancel the divergences found
in the result for the one-loop generating functional of Eq.~(\ref{eq:sl1}).
\par
Up to one loop ${\cal L}_{1}[\psi,J]$ can be written in terms of
a minimal basis of $N$ operators 
$ {\cal O}_i[\psi,J]$. For a non-renormalizable theory, such
as R$\chi$T, $N$
grows with the number of loops. Accordingly we expect to find in our evaluation
of $ S_1[\psi,J]$ many more operators that those in the original
tree level theory $S_0[\psi,J]$. The structure of these
obeys the same construction principles (symmetries) that
gave ${\cal L}_{R\chi T}(\phi,S,P)$ in Eq.~(\ref{eq:lagr1}),
though we foresee  that higher-order chiral tensors may be involved.
A detailed study of the functional integration shows that the new terms
have the structure $\chi^{(4)}$, $ R \, \chi^{(4)} $ or $R \, R \, \chi^{(4)}$
(with a single or multiple traces) and $\chi^{(2)}$, $R\, \chi^{(2)}$ 
and $ R \, R \, \chi^{(2)}$
(with multiple traces)~\footnote{As it will be emphasized later, in the
procedure and due to a necessary field redefinition, terms with more than
two resonances will be generated. We attach to our initial scheme and
only will keep terms with up to two resonances.}.

\subsection{Expansion around the classical solutions}
Following the aforementioned procedure we expand the action associated
to our Lagrangian ${\cal L}_{R \chi T}(\phi,S,P)$ in Eq.~(\ref{eq:lagr1})
around the solutions of the classical EOM~: $u_{cl}(\phi)$, $S_{cl}$
and $P_{cl}$. The fluctuations of the pseudoscalar Goldstone fields 
$\Delta_i$ ($i=0,...,8$), and of the scalar and pseudoscalar resonances 
$\varepsilon_{S_i}$ and 
$\varepsilon_{P_i}$, are parameterized 
as~\footnote{This is a convenient choice for the pseudoscalar fluctuation
variables in order to simplify several cumbersome expressions. Notice
that, once the  ``gauge'' $u_R = u_L^{\dagger} \equiv u$ is enforced,
it implies that the classical and the quantum Goldstone fields 
commute~: $u_{cl} \, \exp(i \Delta /2) = \exp(i \Delta / 2) \, u_{cl}$.}~:
\begin{align} \label{eq:fluc1}
u_R&=\,u_{cl}\,e^{i \Delta / 2}\, ,  &u_L&=\,u_{cl}^\dagger  \, 
e^{-i\Delta / 2}\,, \nonumber \\
S&=\,S_{cl}\,+\,\frac{1}{\sqrt{2}}{\es} \,, 
&P&=\,P_{cl}\,+\,\frac{1}{\sqrt{2}}{\ep}\,,
\end{align}
with 
\begin{align} \label{eq:fluc2}
\Delta &=\,\Delta_i \lam_i / F \,, 
&\es&=\,{\es}_{i}\,\lam_i\,, 
& \ep&=\,{\ep}_{i}\,\lam_i\,.
\end{align}
In the following we will drop the subindex ``$cl$'' for simplicity.
\par
Expanding the Lagrangian using Eqs.~(\ref{eq:fluc1},\ref{eq:fluc2}) up to 
terms quadratic in the fields 
($\Delta_i ,\,\varepsilon_{S_i} ,\,\varepsilon_{P_i}$) and using 
the EOM of Appendix~\ref{app:tres},
we obtain the second-order
fluctuation Lagrangian, that takes the form~\footnote{The intricacies of 
this evaluation are explained in detail in Appendix~\ref{app:cuatro}.}~:
\begin{eqnarray} \label{eq:expansion}
\Delta \mathcal{L}_{\mathrm{R}\chi\mathrm{T}}&=& -\frac{1}{2}\,\Delta_i 
\left( d'_\mu d'^\mu + \sigma \right)_{ij} \Delta_j 
- \frac{1}{2}\,{\es}_{i} \left( d^\mu d_\mu + \ks \right)_{ij} {\es}_{j} 
- \frac{1}{2}\,{\ep}_{i} \left( d^\mu d_\mu + \kp \right)_{ij} {\ep}_{j} 
\nonumber \\
&&+ \,{\es}_{i}\, \as_{ij}\, \Delta_j 
+ \,{\ep}_{i}\, \ap_{ij}\, \Delta_j 
+\,{\ep}_i\, \asp_{ij}\, {\es}_j  \phantom{\frac{1}{2}} \nonumber \\ 
&&
+ \,{\es}_{k}\, \bs_{\mu \, ki}\, d^\mu_{ij} \Delta_j 
+ \,{\ep}_{k}\, \bp_{\mu \, ki}\, d^\mu_{ij} \Delta_j  
+\, {\ep}_k \, \bsp_{\mu \, ki}\, d^\mu_{ij} {\es}_j   \,\, . 
\phantom{\frac{1}{2}} 
\end{eqnarray}
Derivatives and matrices are defined in Appendix~\ref{app:cuatro} where it is 
also shown that in order to write $\Delta \mathcal{L}_{R \chi T}$ 
in the form displayed above we need
to perform two field redefinitions. This procedure generates 
operators with multiple resonance fields. However
our theory, as specified in Section~2, does not include operators 
with more than two resonances and,
for consistency, we shall keep this structure in the fluctuation
 Lagrangian, thus
disregarding operators with three or more resonance
fields in the following. We will comment later on the
 consequences of this feature.
It is customary to write the second-order fluctuation Lagrangian as ~:
\begin{equation} \label{eq:gausso}
\Delta {\cal L}_{\mathrm{R} \chi \mathrm{T}} \, = \, 
- \, \frac{1}{2} \, \eta \, \left( \, \Sigma_{\mu} \, \Sigma^{\mu} \, + \, 
\Lambda \, \right) \, \eta^{\top} \; , 
\end{equation}
where $\eta$ collects the fluctuation fields, 
$\eta=\left(\Delta_i,{\es}_j,{\ep}_k\right)$, $i,j,k = 0,...,8$, 
$\eta^{\top}$ is its transposed and the rest of definitions are given 
in Appendix~\ref{app:cuatro}.

\subsection{Divergent part of the generating functional at one loop}

After we have performed the second-order fluctuation on our Lagrangian
theory we come back to our discussion at the beginning of this Section 
in order to identify the one-loop generating functional, specified now
by the action~:
\begin{equation}
S_{1} \, = \, \frac{i}{2} \, \ln \, \mbox{det} \, 
\left( \, \Sigma_{\mu} \, \Sigma^{\mu} \, + \, \Lambda \, \right) \; .
\end{equation}
We use dimensional regularization to extract the 
divergence of this expression. As emphasized in the literature
\cite{Barvinsky:1985an} it is convenient to employ the Schwinger-DeWitt
proper-time representation, embedded in the heat-kernel formalism,
in order to extract the residue at the $D-4$ pole. Ref.~\cite{bfm} 
shows that, in fact, symmetry considerations can also provide this 
information (at least up to one loop).
\par
Hence we get~:
\begin{equation}\label{eq:oneloop}
S_{1}=-\frac{1}{(4\pi)^2} \, \frac{1}{D-4} 
\int \mathrm{d}^4x \,\, \mbox{Tr}\,  \left( \, \frac{1}{12} \, Y_{\mu\nu}\, 
Y^{\mu\nu} \, + \,  \frac{1}{2} \Lambda^2 \, \right) \, + \, 
S_1^{\mathrm{finite}}\,  , 
\end{equation}
where $\mbox{Tr}$ is short for the trace in the flavour space,
$Y_{\mu \nu}$
denotes the field strength tensor of $Y_{\mu}$ in Eq.~(\ref{eq:y}):
$Y_{\mu \nu} = 
\partial_{ \mu} Y_{\nu} - \partial_{\nu} Y_{\mu} + [Y_{\mu}, Y_{\nu}]$. The 
finite remainder $S_1^{\mathrm{finite}}$ cannot be simply expressed
 as a local Lagrangian, but can be worked out 
for a given transition~\cite{Gasser:1983yg,Unterdorfer:2002zg}.
\par
Finally we get the one-loop divergence as~:
\begin{eqnarray} \label{eq:s1div}
{S}_{1}^{\mathrm{div}}
&= & - \, \frac{1}{(4 \, \pi)^2} \, \frac{1}{D-4} \, \int d^4 x \, 
{\cal L}_{1}^{\mathrm{div}} \; , 
\end{eqnarray}
where
\begin{eqnarray}
{\cal L}_{1}^{\mathrm{div}} & = &   
\,\frac{1}{12} \langle \gamma_{\mu\nu}^{\prime} \gamma^{\prime\,\mu\nu} 
+ 2 \gamma_{\mu\nu}\gamma^{\mu\nu}\rangle \,
+\frac{1}{2} \langle \sigma^2 + \kp^2 + \ks^2 \rangle 
+\langle \as {\as}^\top + \ap {\ap}^\top + \asp {\asp}^\top \rangle
\nonumber \\ 
&&
- \frac{1}{12} \langle  \gamma^{\prime\,\mu\nu} \left( {\bs_{\mu}}^\top 
\bs_{\nu} +{\bp_{\mu}}^\top \bp_{\nu} \right) \rangle \,
 - \, \frac{1}{12} \langle \gamma^{\mu\nu}
\left( \bs_{\mu}{\bs_\nu}^{\top}+\bp_{\mu}{\bp_\nu}^\top+ 
\bsp_{\mu}{\bsp_\nu}^{\top}+{\bsp_{\mu}}^\top \bsp_{\nu} \right)
\rangle  \nonumber \\ 
&&
-\langle {\as}^\top \big( \bar{d}^{\mu}_+ \bs_\mu + \frac{1}{2}
 {\bsp_\mu}^\top \bpm \big) 
 +{\ap}^\top \big( \bar{d}^{\mu}_+ \bp_\mu - \frac{1}{2} {\bsp_\mu} 
 \bsm \big)
+{\asp}^\top \big( \hat{d^\mu} \bsp_\mu + \frac{1}{2} {\bpm} 
{\bs_\mu}^\top \big) \rangle \nonumber \\
&&
+ \frac{1}{4} \langle \sigma \big( {\bs_\mu}^\top \bsm +{\bp_\mu}^\top
 \bpm \big)
+\ks \big(  \bsm {\bs_\mu}^\top+{\bsp_\mu}^\top \bspm \big) 
+ \kp \big(  \bpm {\bp_\mu}^\top+  \bspm{\bsp_\mu}^\top \big) 
\rangle 
\nonumber \\
&&
 +\frac{1}{4} \langle \tilde{d}^{\mu}_- {\bs_\mu}^\top \bar{d}^{\nu}_+ 
\bs_\nu + 
\tilde{d}^{\mu}_- {\bp_\mu}^\top \bar{d}^{\nu}_+ \bp_\nu +
\hat{d^\mu} {\bsp_\mu}^\top \hat{d^\nu} \bsp_\nu \rangle \nonumber \\
&&\
 -\frac{1}{12} \langle \tilde{d}_{+\mu} {\bs_\nu}^\top \bar{d}^{[\mu}_-
 \bsn^{]} + 
\tilde{d}_{+\mu} {\bp_\nu}^\top \bar{d}^{[\mu}_- \bpn^{]} +
\hat{d_\mu} {\bsp_\nu}^\top \hat{d}^{[\mu} \bspn^{]} \rangle \nonumber \\
&&
+\frac{1}{4} \langle \tilde{d}^{\mu}_- {\bs_\mu}^\top {\bsp_\nu}^\top 
\bpn
-\tilde{d}^{\mu}_- {\bp_\mu}^\top \bsp_\nu \bsn
+\hat{d^\mu} {\bsp_\mu}^\top \bpn {\bs_\nu}^\top \rangle \nonumber \\
&&
-\frac{1}{12} \langle \tilde{d}^{\mu}_+ {\bsn}^\top {\bsp_{[\mu}}^\top 
\bp_{{\nu]}}
-\tilde{d}^{\mu}_+ {\bpn}^\top {\bsp_{[\mu}} \bs_{{\nu]}}
+\hat{d^\mu} {\bspn}^\top {\bp_{[\mu}} {\bs_{{\nu]}}}^\top
 \rangle \nonumber \\
&&
+\frac{1}{48} \langle \big( {\bs_\mu}^\top \bsm{\bs_\nu}^\top\bsn+ 
{\bs_\mu}^\top \bsn{\bs_\nu}^\top\bsm+{\bs_\mu}^\top \bs_\nu {\bsm}^\top 
\bsn\big)\nonumber \\
&& \qquad
 +\big( {\bp_\mu}^\top \bpm{\bp_\nu}^\top\bpn+ {\bp_\mu}^\top 
\bpn{\bp_\nu}^\top\bpm+{\bp_\mu}^\top \bp_\nu {\bpm}^\top \bpn\big)\nonumber \\
&& \qquad
+\big( {\bsp_\mu}^\top \bspm{\bsp_\nu}^\top\bspn+ {\bsp_\mu}^\top 
\bspn{\bsp_\nu}^\top\bspm 
\, \, +\,  {\bsp_\mu}^\top \bsp_\nu {\bspm}^\top \bspn\big)
\rangle\nonumber \\
&&
+\frac{1}{24} \langle \big( {\bs_\mu}^\top \bsm{\bp_\nu}^\top\bpn+ 
{\bs_\mu}^\top \bsn{\bp_\nu}^\top\bpm+{\bs_\mu}^\top \bs_\nu {\bpm}^\top 
\bpn\big)\nonumber \\
&& \qquad
+\big( {\bsp_\mu}^\top \bspm{\bsn}{\bs_\nu}^\top+ {\bsp_\mu}^\top
 \bspn{\bs_\nu}{\bsm}^\top+{\bsp_\mu}^\top \bspn {\bsm} {\bs_\nu}^\top 
 \big)\nonumber \\
&& \qquad
 +\big(  \bpm {\bp_\mu}^\top{\bspn}{\bsp_\nu}^\top+ \bpm 
{\bp_\nu}^\top {\bspn}{\bsp_\mu}^\top  +\bp_\mu{\bp_\nu}^\top 
 {\bspm} 
{\bspn}^\top \big)
\rangle \, ,
\end{eqnarray}
where derivatives and matrices are defined in Appendix~\ref{app:cuatro}
and $\gamma_{\mu \nu} = \partial_{\mu} \gamma_{\nu} - 
\partial_{\nu} \gamma_{\mu} + [ \gamma_{\mu}, \gamma_{\nu} ]$
(correspondingly for $\gamma_{\mu \nu}'$).
Moreover for two vectors
$A_{\mu}$,$ B_{\mu}$ we write $A_{[\mu}B_{\nu]}= A_\mu B_\nu - A_\nu B_\mu$.
This result is completely general for the second-order fluctuation Lagrangian
in Eq.~(\ref{eq:expansion}). However, and as explained in Appendix D, the 
expressions given there are valid only for operators with up to two resonances
as we limit ourselves in this article.

\subsection{Result}

When worked out, $S_1^{\mathrm{div}}$ in Eq.~(\ref{eq:s1div}) can be 
expressed in a basis of operators that satisfy the same symmetry requirements
than our starting Lagrangian ${\cal L}_{\mathrm{R} \chi \mathrm{T}}(\phi, S,P)$.
A minimal basis of $R\chi T$ operators that, upon integration of the 
resonances, contributes to the ${\cal O}(p^6)$ $\chi PT$ Lagrangian, in 
$SU(3)$, will be found elsewhere~\cite{nosfuture}. However, up to now,
a basis for the one-loop $R \chi T$ has still not been worked out. This
is precisely our result generated by $S_1^{\mathrm{div}}$. Hence, at one
loop, the $R\chi T$ Lagrangian needed to renormalize our theory reads~:
\begin{equation} \label{eq:main}
{\cal L}_{1} \, = \,   \sum_ {i=1}^{18} \, \alpha_i \, 
{\cal O}_i \, + \, \sum_{i=1}^{68} \, \beta_i^R \, {\cal O}_i^R \, + \, 
\sum_{i=1}^{383} \, \beta_i^{RR} \, {\cal O}_i^{RR} \; .
\end{equation}
The ${\cal O}_i$ operators correspond to those up to ${\cal O}(p^4)$ in 
$U(3)_L \otimes U(3)_R$ $\chi PT$ \cite{Herrera-Siklody:1996pm}. 
${\cal O}_i^{R}$ and ${\cal O}_i^{RR}$ involve one and two resonance
fields, respectively, together with $\chi^{(2)}$ and $\chi^{(4)}$ chiral
tensors. The couplings in the bare Lagrangian $ {\cal L}_{1}$ read,
in accordance with Eq.~(\ref{eq:renoo})~:
\begin{eqnarray} \label{eq:rge}
\alpha_i & = & \mu^{D-4} \, \left( \, \alpha_i^r(\mu) \, + \, 
\frac{1}{(4\pi)^2} \, \frac{1}{D-4} \, \gamma_i  \right) \; , \nonumber \\
\beta_i^R & = & \mu^{D-4} \, \left( \, \beta_i^{R,r}(\mu) \, + \, 
\frac{1}{(4\pi)^2} \, \frac{1}{D-4} \, \gamma_i^R  \right) \; , \nonumber \\
\beta_i^{RR} & = & \mu^{D-4} \, \left( \, \beta_i^{RR,r}(\mu) \, + \, 
\frac{1}{(4\pi)^2} \, \frac{1}{D-4} \, \gamma_i^{RR}  \right) \; ,
\end{eqnarray}
where  $\gamma_i$, $\gamma_i^R$ and $\gamma_i^{RR}$ are the
divergent coefficients given by 
$S_1^{\mathrm{div}}$ that constitute the $\beta$-function of our
Lagrangian (we use the terminology of Ref.~\cite{Buchler:2003vw}).
The determination of the latter
though straightforward involves a long calculation. In order
to diminish the possibility of errors we have performed two independent
evaluations. One of them has been carried out with the help of the 
FORM~3 program \cite{Vermaseren:2000nd} and the other with 
Mathematica \cite{Wolfram}.
The full result when scalar and pseudoscalar resonances
are included is rather lengthy. In Appendix~\ref{app:cinco} we provide
the sets ${\cal O}_i$, ${\cal O}_i^S$ and ${\cal O}_i^{SS}$. The 
coefficients $\gamma_i$ are fully given, however $\gamma_i^S$ and 
$\gamma_i^{SS}$ are only brought when just scalar resonances are considered,
which reduces the total number of operators to 177.
The complete result with the inclusion of pseudoscalar resonances
is available at \texttt{http://ific.uv.es/quiral/rt1loop.html} 
or upon request from the authors.

\section{Features and use of the renormalized \rcht Lagrangian}

In order to understand the aspects and use of the renormalized $R \chi T$
Lagrangian that we have obtained above, we would like to emphasize
here several of its features~:
\begin{itemize}
\item[1/] In Table~\ref{tab:e1} we have collected the full basis
 of ${\cal O}(p^2)$ and
${\cal O}(p^4)$ $U(3)_L \otimes U(3)_R$ $\chi$PT
operators generated in the functional integration at one loop. We 
should recover the result first obtained in Ref.~\cite{Herrera-Siklody:1996pm}.
After the comparison is made~\footnote{Notice that the notation
of Ref.~\cite{Herrera-Siklody:1996pm} is different to ours though, to ease
the comparison, the order chosen is the same. We always
quote our notation for the operators.} 
we agree indeed with their results.
Notice though that in order to disentangle the resonances,
it is not enough to withdraw
all the resonance couplings. This is because the derivative terms in
${\cal L}_{\mathrm{kin}}(S,P)$, which do not carry any resonance coupling,
also contribute through the functional
integration to several of the operators, namely 
${\cal O}_4$, ${\cal O}_7$, ${\cal O}_{13}$, ${\cal O}_{14}$
and ${\cal O}_{15}$
in Table~\ref{tab:e1}. We have confirmed that 
${\cal L}_{\mathrm{kin}}(S,P)$ gives precisely the difference between
our coefficients
$\gamma_4$, $\gamma_7$, $\gamma_{13}$, $\gamma_{14}$, $\gamma_{15}$  
and those of Ref.~\cite{Herrera-Siklody:1996pm} 
once the resonance couplings have been switched off.
\item[2/] Our result provides the running of the $\alpha_i$, 
$\beta_i^R$ and $\beta_i^{RR}$ couplings through the renormalization
group equations (RGE). From Eq.~(\ref{eq:rge}) we get~:
\begin{equation} \label{eq:rgeq}
\mu \, \frac{d}{d \mu} \, \alpha_i^r(\mu) \, = \, 
- \, \frac{\gamma_i}{16 \,\pi^2} \; , 
\end{equation}
and, analogously, for $\beta_i^R$ and $\beta_i^{RR}$. This result can be
potentially useful if we are interested in the phenomenological evaluation
of the resonance couplings at this order. Though $\mu$ is known
to be of the order of a typical scale of the physical system, let us say
$\mu = M_S$ or $\mu = M_P$, there always remains some ambiguity on the 
precise value of $\mu$ at which the low-energy couplings are extracted
from the phenomenology.
The RGE (\ref{eq:rgeq}) provides an estimate of the reliance of 
such determinations.
If the coupling under request varies drastically with
the scale it is clear that the value obtained phenomenologically
has a large uncertainty, while if it has a smooth running 
the determination is more reliable. 
\par
Within this issue it is interesting to take a closer look to the 
running of the resonance couplings in the original \rcht 
Lagrangian \cite{Ecker:1988te}, namely, $c_d^r(\mu)$, $c_m^r(\mu)$ and
$d_m^r(\mu)$. Their corresponding $\beta$-function coefficients are
$\beta_1^S$, $\beta_4^S$ and $\beta_{44}^P$, respectively (notice that we are
taking into account here the complete results, including scalar and
pseudoscalar resonances). We obtain~:
\begin{eqnarray} \label{eq:evolution}
\mu \, \frac{d}{d  \mu} \, c_d^r (\mu) & = & - \, \frac{N}{16 \pi^2} \, 
c_d \, \frac{M_S^2}{F^2} \, \left[ \, 2 \lambda_1^{SS} \, + \, 
4 \lambda_2^{SS} \, - \, 3 \frac{c_d^2}{F^2} \, - \, 
 (\lambda_1^{SP})^2 \, \right] \, , \nonumber \\ & & \nonumber \\
\mu \, \frac{d}{d  \mu} \, c_m^r (\mu) & = & - \,\frac{N}{32 \pi^2} \, 
\frac{M_S^2}{F^2} \, \left[ \, c_d \left( \, 1 \, - \, 4 \, 
\frac{c_d c_m}{F^2} \right) \,
- \, 4 \,  \lambda_1^{SP} \, \left( \, d_m \, + \,  c_m
 \, \lambda_1^{SP} \, 
\right) \, \right] \, ,   \\  & & \nonumber  \\
\mu \, \frac{d}{d  \mu} \, d_m^r (\mu) & = & - \, \frac{N}{16 \pi^2} \,
\lambda_1^{SP} \, \frac{M_P^2}{F^2}  
\left(   1  -   2  \frac{M_S^2}{M_P^2} \right) \, 
\left[  c_d \,  
- \, 2 \,  c_m \, - \, 
2 \, d_m \,  
\lambda_1^{SP}  \right] 
\, ,
\nonumber
\end{eqnarray}
where $N$ is the number of flavours. First of all note that
the running, as expected, is a next-to-leading order effect in the 
$1/N_C$ expansion. This fact is guided by the $M_R^2/F^2$ factors 
taking into account that $M_R \sim {\cal O}(1)$ and 
$F \sim {\cal O}(\sqrt{N_C})$.
Another interesting aspect is the interval over which $\mu$ runs. It is 
well known \cite{Collins} that the couplings are only relevant at the 
scale of the momenta involved in the processes (in order to diminish the
role of the logarithms). In our case $\mu \sim M_S, \, M_P$. Thus we do not
expect a large running for the scale, namely a few hundreds of MeV. This
last conclusion brings us to the next point. At next to leading
order in $1/N_C$ we can
ignore the running on the right-hand-side of Eqs.~(\ref{eq:evolution}). 
Hence we can input the leading order values for the couplings, given
by Eqs.~(\ref{eq:largecd},\ref{eq:largecd2}) to obtain the leading 
logarithm in the evolution of $c_m^r(\mu)$ and $d_m^r(\mu)$. It is remarkable
that, at this order, Eqs.~(\ref{eq:evolution}) predict no running 
for these coupling constants, as the right-hand-side
of their RGE vanishes. Unfortunately we cannot
conclude anything about $c_d$, as there are no known constraints on
$\lambda_1^{SS}$ and $\lambda_2^{SS}$. If the Large-$N_C$ estimates for 
the couplings are to be reliable we come to the conclusion that the 
predictions for $c_m$ and $d_m$ are rather robust. Moreover the same
exercise for the ${\cal O}_i$ operators involving only Goldstone and
external fields in Tab.~\ref{tab:e1} shows that also 
${\cal O}_2$, ${\cal O}_8$, ${\cal O}_9$, ${\cal O}_{10}$, ${\cal O}_{11}$,
${\cal O}_{12}$, ${\cal O}_{16}$ and ${\cal O}_{18}$ do not run at one loop,
i.e. all those operators involving $\chi_+$ and/or $\chi_{-}$. Also notice
that $c_m$ and $d_m$ rule operators sharing this feature.
\item[3/] In the procedure we have employed to evaluate the functional
integration of ${\cal L}_{\mathrm{R} \chi \mathrm{T}}$ up to one loop
we have withdrawn those operators with three or more resonance fields
and kept up to two resonances. A cut in the number of resonances is 
necessary because to reach the Gaussian expression in Eq.~(\ref{eq:gausso})
we need to perform several field transformations 
(see Appendix~\ref{app:cuatro}) that generate operators with 
more resonance fields which in turn 
require additional field transformations and so on. 
One of the differences of \rcht
with respect to \chpt (in the strong \cite{Gasser:1983yg} or 
electroweak interaction \cite{Kambor:1989tz} form of the latter)
is that we 
do not have an expansion parameter into the Lagrangian that can provide
a natural cut for higher order terms in these field transformations.
Notice that the cut in the number of resonances seems to hinder
our result, as it does not allow us to renormalize divergent one loop
diagrams with three or more resonance fields as external legs. However
we would not expect to treat these loops as we are not including, in our 
leading order Lagrangian, interacting terms with three or more resonance
fields.
\end{itemize}
\begin{figure}
\begin{center}
\includegraphics[scale=0.70]{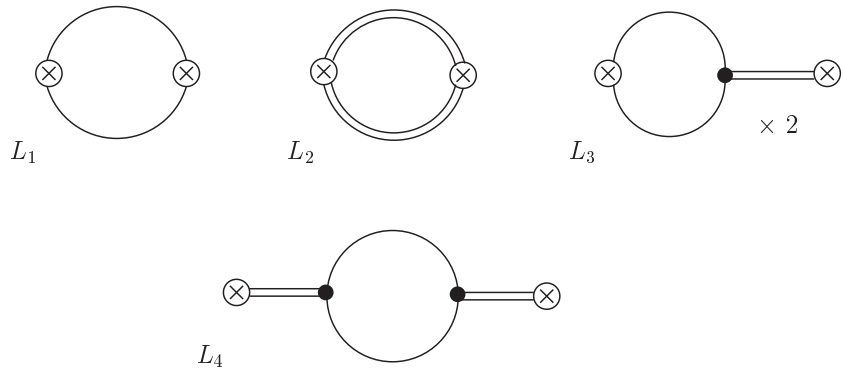}
\caption{\label{fig:loop}
One-loop contributions to the 
$\Pi_{SS}^{ij}(q^2)$ correlator in
the chiral limit when only scalar resonances are included. 
A single line stands for a Goldstone boson while
a double line indicates a scalar resonance. Their result is divergent.}
\end{center}
\end{figure}
To end this Section we would like to show a simple example of the application
of our result. We consider the one-loop renormalization of the two-point
function of scalar currents~:
\begin{equation}
\Pi_{SS}^{ij} (q^2) \, = \, i \, \int  d^4 x \, e^{i q \cdot x} \, 
\langle 0 |T\{ S^i(x) S^j(0)\} | 0 \rangle \, , \qquad \, \;  \; 
S^i(x) = \overline{q}(x) \lambda^i q(x) \; , 
\end{equation}
in the chiral limit and when only scalar resonances are considered. The 
divergent loop diagrams contributing are those depicted in Fig.~\ref{fig:loop}.
In order to cancel the divergences one needs to add the counterterm 
contributions in Fig.~\ref{fig:counter}, where diagram $C_1$ is given by
${\cal O}_{12} + 2 {\cal O}_{16} = \langle \chi_+^2 \rangle$
in Table~\ref{tab:e1}, $C_2$ by 
${\cal O}_{4}^{S}=\langle S \chi_{+} \rangle $ in 
Table~\ref{tab:e2} and $C_3$ by 
${\cal O}_{1}^{SS}= \langle S S \rangle$ in Table~\ref{tab:e3}.
The cancellation works as follows:
One part of the contribution of $C_1$ cancels completely the divergence in the 
loops $L_1+L_2$. Another piece of $C_1$ together with $C_2$ eliminates the 
divergence coming from $L_3$ and, finally, all remaining contributions of 
$C_1$ and $C_2$ add to $C_3$ in order to render $L_4$ finite. Notice that,
as there are no nonlocal divergences, 
the contributions of 1PR diagrams are brought finite once 1PI diagrams
have been properly renormalized.
\begin{figure}[h]
\begin{center}
\vspace*{1.4cm}
\includegraphics[scale=0.68]{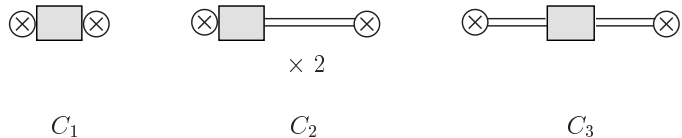}
\caption{\label{fig:counter}
Counterterm contributions that renormalize the one-loop result of 
Fig.~\ref{fig:loop}. A double line stands for a scalar resonance.}
\end{center}
\vspace*{-0.4cm}
\end{figure}

\section{Summary}

\rcht provides a consistent framework to study the energy region of 
the hadronic resonances, $M_V \lsim E \lsim 2 \, \gev$. It embodies a 
phenomenological Lagrangian where Goldstone bosons and resonances fields
are kept as active degrees of freedom; this is the key 
ingredient for the application of the large-$N_C$ expansion. Recently, 
and after its multiple explorations at tree level, it has emerged
some interest in the application of \rcht at one loop level mainly 
to understand how the features of QCD are implemented into the theory.
\par
In this
work we have systematically obtained, by using the background field method
and for the first time, both the full basis of operators and
the $\beta$-function coefficients that render finite, up to one loop, our
initial Lagrangian ${\cal L}_{\mathrm{R} \chi \mathrm{T}}$ in 
Eq.~(\ref{eq:lagr1}). This would correspond to the next-to-leading order in 
the $1/N_C$ expansion but including one multiplet of scalar and pseudoscalar
resonances only. Our main result is given by Eq.~(\ref{eq:main}) and the
$\gamma_i$, $\gamma_i^R$, $\gamma_i^{RR}$ parameters in Eq.~(\ref{eq:rge}).
The outcome is relevant for the study of those diagrams involving a loop
with up to two resonances and any number of Goldstone bosons in the legs.
\vspace*{0.4cm} \\
\noindent{\large\bf Acknowledgements}
\vspace*{0.3cm} \\
The authors would like to thank G.~Amor\'os, G.~Ecker, R.~Kaiser and A.~Pich 
for hints and conversations on the topic of this work. 
P.~Ruiz-Femen\'\i a is indebted to
M.~Aguado and T.~Hahn for their help with Mathematica programming.
I.~Rosell is supported by a FPU scholarship of the Spanish MEC.
This work has been supported in part by
MEC (Spain) under Grant FPA2004-00996, by Generalitat Valenciana
(Grants GRUPOS03/013, GV04B-594 and GV05/015) and by ERDF funds from the 
European Commission.
\appendix
\newcounter{erasmo}
\renewcommand{\thesection}{\Alph{erasmo}}
\renewcommand{\theequation}{\Alph{erasmo}.\arabic{equation}}
\renewcommand{\thetable}{\Alph{erasmo}}
\setcounter{erasmo}{1}
\setcounter{equation}{0}
\setcounter{table}{0}

\section{Notation}
\label{app:uno}

In this article we follow the notation introduced in 
Ref.~\cite{Ecker:1988te} that we collect here.
\par
The non-linear realization of spontaneously broken chiral symmetry
$G = U(3)_L \otimes U(3)_R$ is defined by the action of the group on the
elements $u_{R,L}(\phi)$ of the coset space $G \, / \, U(3)_V$~:
\begin{eqnarray}
u_L(\phi)  & \mapright{G}{} &  g_L \, u_L(\phi) \, h(g,\phi)^{\dagger} \, ,
\nonumber \\
u_R(\phi) & \mapright{G}{} & g_R \, u_R(\phi) \,   h(g,\phi)^{\dagger} \, ,  
\end{eqnarray}
where $g=(g_L,g_R) \in G$ and $h(g,\phi) \in U(3)_V$. 
\par
The following chiral tensors have been used:
\begin{eqnarray}
u_\mu&=&i\,\left\{u_R^\dagger \left(\partial_\mu - i r_\mu\right) 
u_R -u_L^{\dagger} \left(\partial_\mu - i \ell_\mu \right) u_L\right\}\,,
 \nonumber \\ 
\Gamma_\mu &=& \frac{1}{2} \, 
\left\{u_R^\dagger \left(\partial_\mu - i r_\mu
\right) u_R +u_L^{\dagger}\left(\partial_\mu - i\ell_\mu \right) u_L
\right\}\, , 
\nonumber \\
\chi_\pm&=&u_R^\dagger\chi u_L \pm u_L^{\dagger}\chi^\dagger u_R \,,
 \nonumber \\ 
\chi&=& 2B_0\left(s+ip\right) \, , 
\end{eqnarray}
where $s,p,r$ and $\ell$ are scalar, pseudoscalar, right and left external
fields, respectively. The covariant derivative is defined by
\begin{eqnarray}
\nabla_\mu X &=& \partial_\mu X + \left[\Gamma_\mu, X \right] \, ,
\end{eqnarray}
for any $X$ transforming as 
\begin{eqnarray}
X \, \mapright{G}{} \, h(g,\phi) \, X \, h(g,\phi)^ {\dagger}  \; ,
\end{eqnarray}
under the chiral group, like the tensors $u_{\mu}$, $\chi_{\pm}$ or
the resonance fields $S,P$.
\par 
We can take the choice of coset representative such that 
$u_R(\phi) = u_L^{\dagger}(\phi) \equiv u(\phi)$,
whose explicit form in the exponential
parameterization is~:
\begin{eqnarray}
u(\phi) \, = \, \exp \left( \frac{i}{\sqrt{2} \, F} \, \phi \right) \; \; 
& , & \; \; \phi \, = \, \frac{1}{\sqrt{2}} \, \sum_{i=0}^{8} \, \lambda_i \,
\phi_i \; , 
\end{eqnarray}
where the normalization of the Gell-Mann matrices is given by 
$\langle \lambda_i \lambda_j \rangle = 2 \delta_{ij}$ and $\phi$
is the 
nonet of pseudo-Goldstone fields. Scalar and pseudoscalar resonances have
analogous $U(3)_V$ content to the one described by $\phi$~:
\begin{equation}
\phi =  
\left(
\begin{array}{ccc}
 \frac{1}{\sqrt 2}\pi^0 + \frac{1}{\sqrt 6}\eta_8
 + \frac{1}{\sqrt 3}\eta_0 
& \pi^+ & K^+ \\
\pi^- & - \frac{1}{\sqrt 2}\pi^0 + \frac{1}{\sqrt 6}\eta_8 
 + \frac{1}{\sqrt 3}\eta_0  
& K^0 \\
 K^- & \bar{K}^0 & - \frac{2}{\sqrt 6}\eta_8 
 + \frac{1}{\sqrt 3}\eta_0  
\end{array}
\right)
\ . 
\end{equation}

\appendix
\newcounter{aristocles}
\renewcommand{\thesection}{\Alph{aristocles}}
\renewcommand{\theequation}{\Alph{aristocles}.\arabic{equation}}
\renewcommand{\thetable}{\Alph{aristocles}}
\setcounter{aristocles}{2}
\setcounter{equation}{0}
\setcounter{table}{0}

\section{Short-distance constraints on the $\lambda_i^{RR}$ couplings}
\label{app:dos}

Most of the short-distance constraints on the couplings of \rcht come from
matching Green functions of QCD currents evaluated within the resonance
theory with the results obtained in the leading OPE expansion 
\cite{Cirigliano:2004ue}. Another source of conditions arise from form factors.
Parton dynamics demands that two-body form factors of hadronic currents
should vanish at 
infinite momentum transfer \cite{Lepage:1979zb}, property known as 
Brodsky-Lepage behaviour. It is also phenomenologically well known
that any hadronic form factor seems to share this feature, 
at least in the case of stable asymptotic states.
\par
Form factors of QCD currents that involve resonance asymptotic states
do not necessarily fall in the group above. Hence it is not clear whether
the latter should satisfy the Brodsky-Lepage behaviour. 
It is often
claimed that as resonances are non-decaying states in the 
$N_C \rightarrow \infty$ limit, 
they should not be treated differently from (real) stable states. However
one shall keep in mind that the strict large-$N_C$ limit also requires
to consider 
an infinite number of resonances, so the discussion is far from settled.
\par
Leaving aside this issue, if one enforces the Brodsky-Lepage behaviour 
on the resonance form factors 
useful constraints among the couplings are found~\cite{nosfuture}. From the
pseudoscalar form factor $\langle P^i | p^j | S^k \rangle$ and
$\langle P^i | s^j | \pi^k \rangle$ we obtain, respectively,
\begin{equation}
\lambda_1^{SP} \, = \, 4 \, \lambda_2^{SP} \, ,
\end{equation}
and
\begin{equation}
\lambda_1^{SP} \, = \, - \, \frac{d_m}{c_m} \; .
\end{equation}
Finally the scalar form factors $\langle P^i | s^j | P^k \rangle$ and
$\langle S^i | s^j | S^k \rangle$ give
\begin{equation}
\lambda_3^{PP} \, = \, \lambda_3^{SS} \, = \, 0 \; .
\end{equation}

\appendix
\newcounter{ciceron}
\renewcommand{\thesection}{\Alph{ciceron}}
\renewcommand{\theequation}{\Alph{ciceron}.\arabic{equation}}
\renewcommand{\thetable}{\Alph{ciceron}}
\setcounter{ciceron}{3}
\setcounter{equation}{0}
\setcounter{table}{0}

\section{Equations of Motion for the classical fields}
\label{app:tres}

The classical fields are defined by Eq.~(\ref{eq:eom}). 
From the \rcht Lagrangian in Eq.~(\ref{eq:lagr1}),
the EOM's for the Goldstone and resonance fields are obtained as the system
of coupled equations~:
\begin{align}
\nabla^\mu u_\mu \,=&\, \frac{i}{2}  \chi_- 
-\frac{2c_d}{F^2} \nabla^\mu \left\{u_\mu ,S \right\} 
+\frac{i\,c_m}{F^2}  \left\{\chi_- ,S\right\}  \,-\frac{1}{2F^2} 
\big[u_\mu,[\nabla^\mu S,S] \big]
-\frac{2 \ssa}{F^2} \nabla_\mu \left\{ u^\mu, SS\right\} \nonumber \\
& -\frac{4\ssb}{F^2} \nabla_\mu \left( S\, u^\mu S\right)\,+\,
\frac{i\,\ssc}{F^2} \left\{ \chi_- ,SS\right\}
-\frac{d_m}{F^2}  \left\{\chi_+ , P\right\} \,-\,\frac{1}{2F^2} 
\big[u_\mu,[\nabla^\mu P,P] \big] \nonumber \\
&-\,\frac{2 \ppa}{F^2} \nabla_\mu \left\{ u^\mu, PP\right\} 
-\,\frac{4\ppb}{F^2} \nabla_\mu \left( P\, u^\mu P\right)\,
+\,\frac{i\,\ppc}{F^2} \left\{ \chi_- ,PP\right\}
-\frac{2\spa}{F^2} \nabla^\mu \{ \nabla_\mu S, P \} \nonumber \\
& +\frac{\spa}{2F^2} \Big[ u_\mu , \big[ S , \{ P, u^\mu \} \big] \Big]  
-\frac{\spb}{F^2} \big\{ \chi_+ , \{ S, P \} \big\} \, , \\
& & \nonumber 
\end{align}
\begin{align}
\nabla^\mu \nabla_\mu S \,=&\, -M_S^2 \,S \,+\, c_m \, \chi_+ 
 \,+\,c_d \, u_\mu u^\mu  \, 
+\, \ssa \left\{ S, u_\mu u^\mu \right\}\,+\, 2 \ssb u_\mu S u^\mu 
\phantom{\frac{1}{2}}\nonumber \\
&+\, \ssc \left\{ S, \chi_+ \right\}\,  -\, \lam_1^{\mathrm{SP}} 
\nabla_\mu \{ P , u^\mu \}
\,+\, i \lam_2^{\mathrm{SP}} \{ P , \chi_- \}  \, , \phantom{\frac{1}{2}}\\
& & \nonumber \\
\nabla^\mu \nabla_\mu P \,=&\, -M_P^2 \,P \,+\, i \,d_m \, \chi_- \,+\,
 \ppa \left\{ P, u_\mu u^\mu \right\}\,+\, 2 \ppb u_\mu P u^\mu 
 \phantom{\frac{1}{2}} \nonumber \\
& +\, \ppc \left\{ P, \chi_+ \right\} \, +\, \lam_1^{\mathrm{SP}}  
\{ \nabla_\mu S , u^\mu \}
\,+\, i \lam_2^{\mathrm{SP}} \{ S , \chi_- \}\, \,.\phantom{\frac{1}{2}}
\end{align}

\appendix
\newcounter{catilina}
\renewcommand{\thesection}{\Alph{catilina}}
\renewcommand{\theequation}{\Alph{catilina}.\arabic{equation}}
\renewcommand{\thetable}{\Alph{catilina}}
\setcounter{catilina}{4}
\setcounter{equation}{0}
\setcounter{table}{0}

\section{Second-order fluctuation of the Lagrangian}
\label{app:cuatro}

The expansion around the classical solution of the fields in our
Lagrangian ${\cal L}_{R \chi T}(\phi,S,P)$ up to second order
(as required for the one loop evaluation) gives~:
\begin{eqnarray}
\Delta \mathcal{L}_{\mathrm{R}\chi\mathrm{T}}
&=&\Delta \mathcal{L}_{\chi}^{(2)}\,+\,\Delta \mathcal{L}_{\mathrm{kin}}
(\mathrm{S},\mathrm{P}) 
\,+\,  \Delta \mathcal{L}_{2}(\mathrm{S})\, +\, 
 \Delta \mathcal{L}_{2}(\mathrm{P})\,+\, 
 \Delta \mathcal{L}_{2}(\mathrm{S},\mathrm{P}) \, ,\quad 
\end{eqnarray}
where
\begin{align}
\Delta \mathcal{L}_{\chi}^{(2)}\,=&-\frac{F^2}{8}\bra 
\chi_+ \Delta^2 \ket \,+\, \frac{F^2}{4} \bra \nabla^\mu \Delta
 \nabla_\mu \Delta \,+\, \frac{1}{4} \left[ u_\mu,\Delta \right] 
 \left[ u^\mu,\Delta \right] \ket \,, 
\end{align}
\begin{align}
\Delta \mathcal{L}_{\mathrm{kin}}(\mathrm{S},\mathrm{P})\,=
&\,\frac{1}{4}\bra \nabla^\mu \es \nabla_\mu \es \ket 
\,-\, \frac{M_S^2}{4}\bra \es \, \es \ket \,+\, \frac{1}{32} \bra 
\big[ [u^\mu,\Delta],S\big] \big[ [u_\mu,\Delta],S\big] \ket \,\nonumber \\ 
& -\, \frac{1}{8} \bra  [\nabla_\mu \Delta,\Delta][S, \nabla^\mu S]
\ket \,+\, \frac{1}{4\sqrt{2}} \bra  [ u_\mu,\Delta]
\Big( [S, \nabla^\mu \es]-[ \nabla^\mu S, \es]\Big) \ket \nonumber \\ 
& +\,\frac{1}{4}\bra 
\nabla^\mu \ep \nabla_\mu \ep \ket \,-\, \frac{M_P^2}{4}\bra \ep \, 
\ep \ket \,+\, \frac{1}{32} \bra \big[ [u^\mu,\Delta],P\big] 
\big[ [u_\mu,\Delta],P\big] \ket \,\nonumber \\ 
& -\, \frac{1}{8} \bra  [\nabla_\mu \Delta,\Delta][P, \nabla^\mu P]
\ket \,+\, \frac{1}{4\sqrt{2}} \bra  [ u_\mu,\Delta] 
\Big( [P, \nabla^\mu \ep] - [\nabla^\mu P, \ep] \Big) \ket \,,
\end{align}
\begin{align}
\Delta \mathcal{L}_{2}(\mathrm{S})\,=&-\frac{i\,c_m}{2\sqrt{2}}\bra 
\es \{\Delta,\chi_-\} \ket \,-\, \frac{c_m}{8} \bra \{S, \Delta\} 
\{\chi_+,\Delta\}\ket - \frac{c_d}{\sqrt{2}} \bra \es \{\nabla_\mu 
\Delta, u^\mu \} \ket\nonumber \\
&\, + \bra \left( c_d \, S + \ssa SS \right) \left( \nabla^\mu 
\Delta \nabla_\mu \Delta +\frac{1}{8} \Big \{  \big[\Delta,[u_\mu,\Delta]
\big],u^\mu\Big\} \right)  \ket \, 
+\,\frac{\ssa}{2} \bra \varepsilon^{2}_{\mathrm{S}}\, u^\mu u_\mu \ket 
 \nonumber \\
&-\, \frac{\ssa}{\sqrt{2}} \bra \left\{ S, \es \right\} \left\{ u_\mu , 
\nabla^\mu \Delta\right\} \ket
\,+\, \ssb \bra S\, \nabla_\mu \Delta\, S \,\nabla^\mu \Delta \ket \,+\,
\frac{\ssb}{2}\bra \es u_\mu \es u^\mu \ket \nonumber \\
&-\sqrt{2}\ssb \bra \es \big( \nabla_\mu \Delta \,S \,u^\mu + u_\mu S \,
\nabla^\mu \Delta \big) \ket  \,+\,
\frac{\ssb}{4} \bra  \big[ [\Delta, u_\mu ],\Delta \big] S \,u^\mu S \ket 
 \nonumber \\&
-\,\frac{\ssc}{8} \bra \{ SS ,\Delta \}  \{\chi_+,\Delta\}  \ket \,-\,
 \frac{i\, \ssc}{2\sqrt{2}}  \bra \{ S, \es \} \{ \chi_- , \Delta \} \ket
\,+\,  \frac{\ssc}{2} \bra \varepsilon^{2}_{\mathrm{S}} \,\chi_+ \ket \,,
\label{problem}\\ \nonumber 
\end{align}
\begin{align}
\Delta \mathcal{L}_{2}(\mathrm{P})\,=&\,\frac{d_m}{2\sqrt{2}}\bra \ep 
\{\Delta,\chi_+\} \ket \,-\, \frac{i\,d_m}{8} \bra \{ P, \Delta\} 
\{\chi_-,\Delta\}\ket  \nonumber \\
&\,+\ppa \bra PP \left( \nabla^\mu \Delta \nabla_\mu \Delta +\frac{1}{8} 
\Big\{  \big[\Delta,[u_\mu,\Delta]\big],u^\mu\Big\} \right)  \ket  
\,+\,\frac{\ppa}{2} \bra \varepsilon^{2}_{\mathrm{P}}\, u^\mu u_\mu \ket
\nonumber \\
& -\, \frac{\ppa}{\sqrt{2}} \bra \left\{ P, \ep \right\} \left\{ u_\mu ,
 \nabla^\mu \Delta\right\} \ket
\,+\, \ppb \bra P\, \nabla_\mu \Delta\, P \,\nabla^\mu \Delta \ket \,+\,
\frac{\ppb}{2}\bra \ep u_\mu \ep u^\mu \ket \nonumber \\
&-\sqrt{2}\ppb \bra \ep \big( \nabla_\mu \Delta \,P \,u^\mu + u_\mu P \,
\nabla^\mu \Delta \big) \ket\,   \ket \,+\,
\frac{\ppb}{4} \bra  \big[ [\Delta, u_\mu ],\Delta \big] P \,u^\mu P \ket  
\nonumber \\&
-\,\frac{\ppc}{8} \bra \{ PP ,\Delta \}  \{\chi_+,\Delta\}  \ket \,-\,
 \frac{i\, \ppc}{2\sqrt{2}}  \bra \{ P, \ep \} \{ \chi_- , \Delta \} \ket
\,+\,  \frac{\ppc}{2} \bra \varepsilon^{2}_{\mathrm{P}} \,\chi_+ \ket \,, 
\label{problem2} 
\end{align}
\begin{align}
\Delta \mathcal{L}_{2}(\mathrm{S},\mathrm{P})\,=&\frac{\spa}{8}\bra 
\{ \nabla_\mu S, P \} \big[ [\Delta, u^\mu ], \Delta \big] \ket
-\frac{\spa}{\sqrt{2}} \bra  \nabla^\mu \Delta \big( \{ \nabla_\mu \es ,
 P\} + \{ \nabla_\mu S,\ep\}  \big)\ket \nonumber \\
&+  \frac{\spa}{4\sqrt{2}} \bra \big[ [ u_\mu, \Delta ], S  \big] \big( 
\{\ep , u^\mu \} - \sqrt{2}\{ P , \nabla^\mu \Delta \} \big) \ket
+\frac{\spa}{2} \bra \{\nabla_\mu \es, \ep \} u^\mu \ket \nonumber \\
& +\frac{\spa}{4\sqrt{2}} \bra \big[ [u_\mu, \Delta ], \es \big] \{ P , 
u^\mu\} \ket
+\frac{\spa}{8} \bra \big[ [ \Delta, \nabla_\mu \Delta ] , S \big] \{ P , 
u^\mu\} \ket \nonumber \\
&-\frac{i\, \spb}{8} \bra \{ S,P \} \big\{\Delta, \{ \chi_- , \Delta \}
 \big\} \ket
+\frac{\spb}{2\sqrt{2}} \bra \{ \Delta, \chi_+ \} \Big( \{\es, P \} +
 \{ S, \ep\} \Big) \ket \nonumber \\
& +\frac{ i\, \spb}{2} \bra \chi_- \{ \es, \ep \} \ket   \, .\label{problem3} 
\end{align}

The evaluation of the path integral requires a Gaussian rearrangement of
the integration variables.
However the second-order fluctuation
$\Delta {\cal L}_{R \chi T}$ does not have this structure due to the terms
$\bra PP \,\nabla_\mu \Delta \nabla^\mu \Delta \ket$,
$\bra P \,\nabla_\mu  \Delta\,P\, \nabla^\mu \Delta \ket$, 
$ \bra S \nabla_\mu \Delta \nabla^\mu \Delta \ket $,
$\bra SS \,\nabla_\mu \Delta \nabla^\mu \Delta \ket$,
$\bra S \,\nabla_\mu  \Delta\,S\, \nabla^\mu \Delta \ket$
and $\bra \{ \nabla_\mu \es, P \} \nabla^\mu \Delta \ket $
in Eqs.(\ref{problem},\ref{problem2},\ref{problem3}). A way out is provided
by a redefinition of the fields that eliminates the unwanted terms~:
\begin{align} \label{eq:redefine}
\Delta \, \rightarrow \, & \, \Delta  \,- \frac{c_d}{F^2} 
\left\{ \Delta, S \right\} 
- \frac{\tilde{\lam}^{\mathrm{SS}}_1}{F^2} \left\{\Delta ,SS\right\} -
\frac{2\tilde{\lam}^{\mathrm{SS}}_2}{F^2} S\, \Delta \, S \, 
- \frac{\tilde{\lam}^{\mathrm{PP}}_1}{F^2} \left\{\Delta ,PP\right\} -
\frac{2\tilde{\lam}^{\mathrm{PP}}_2}{F^2} P\, \Delta \, P \,, \nonumber \\
\es \, \rightarrow \, & \, \es \,+ \sqrt{2} \spa \{ P, \Delta\} - 
\frac{\sqrt{2}\spa c_d}{F^2} \big\{ P , \{ \Delta, S \} \big\} \, ,
\end{align} 
where the following constants have been defined~:
\begin{align}
\tilde{\lam}^{\mathrm{SS}}_1&\equiv \, \lam^{\mathrm{SS}}_1 
-\frac{3}{2}\frac{c_d^2}{F^2} \,, &
\tilde{\lam}^{\mathrm{SS}}_2&\equiv \, \lam^{\mathrm{SS}}_2 
-\frac{3}{2}\frac{c_d^2}{F^2} \,, \nonumber \\
\tilde{\lam}^{\mathrm{PP}}_1&\equiv \, \lam^{\mathrm{PP}}_1 
-{(\spa)}^2 \,, \phantom{\frac{1}{2}} &
\tilde{\lam}^{\mathrm{PP}}_2&\equiv \, \lam^{\mathrm{PP}}_2 
-{(\spa)}^2 \,.\phantom{\frac{1}{2}}
\end{align}
The transformation of the integration measure only yields $\delta^4(0)$
terms which
have no effect on the theory~\cite{Politzer:1980me}~\footnote{In dimensional
regularization the later result is immediate, as $\delta^d(0)=0$.}.

Performing the transformations given by Eq.~(\ref{eq:redefine}) on
$\Delta \mathcal{L}_{\mathrm{R}\chi\mathrm{T}}$
and keeping only terms with up to two resonances we finally obtain~:
\begin{eqnarray} \label{eq:rchtapp}
\Delta \mathcal{L}_{\mathrm{R}\chi\mathrm{T}}&=& -\frac{1}{2}\,\Delta_i 
\left( d'_\mu d'^\mu + \sigma \right)_{ij} \Delta_j 
- \frac{1}{2}\,{\es}_{i} \left( d^\mu d_\mu + \ks \right)_{ij} {\es}_{j} 
- \frac{1}{2}\,{\ep}_{i} \left( d^\mu d_\mu + \kp \right)_{ij} {\ep}_{j} 
\nonumber \\
&&+ \,{\es}_{i}\, \as_{ij}\, \Delta_j 
+ \,{\ep}_{i}\, \ap_{ij}\, \Delta_j 
+\,{\ep}_i\, \asp_{ij}\, {\es}_j  \phantom{\frac{1}{2}} \nonumber \\ 
&&
+ \,{\es}_{k}\, \bs_{\mu \, ki}\, d^\mu_{ij} \Delta_j 
+ \,{\ep}_{k}\, \bp_{\mu \, ki}\, d^\mu_{ij} \Delta_j  
+\, {\ep}_k \, \bsp_{\mu \, ki}\, d^\mu_{ij} {\es}_j   \,\, , 
\phantom{\frac{1}{2}} 
\end{eqnarray}
that has the proper Gaussian structure and where the
following definitions have been introduced~:

\begin{eqnarray}
d^\mu_{ij}&=&\delta_{ij}\,\partial^\mu + {\gamma_{ij}^\mu}\big|_{\chi}
 \, ,\\ \nonumber \\ 
d'^\mu_{ij}&=& d^\mu_{ij} +{\gamma_{ij}^\mu}\big|_{\mathrm{R}} \, , \\
\nonumber \\
{\gamma_{ij}^\mu}\big|_{\chi}&=&-\frac{1}{2} \bra \Gamma^\mu 
[\lam_i,\lam_j] \ket \, , \\ \nonumber \\
{\gamma_{ij}^\mu}\big|_{\mathrm{R}}&=&\frac{c_d \spa}{2F^2}
 \bra \{ P, \lam_i\} \{ u^\mu, \lam_j \} \ket 
+ (-\frac{1}{16F^2}+\frac{c_d^2}{8F^4}) \bra [S,\nabla^\mu S]\, 
[\lam_i,\lam_j] \ket   \nonumber \\
&& -\frac{1}{16F^2} \bra [P,\nabla^\mu P]\, [\lam_i,\lam_j] \ket    
-\frac{\spa}{16F^2} \bra \big[ S, \{ P, u^\mu \} \big] [\lam_i, 
\lam_j ] \ket \nonumber \\
&& +\frac{\ssb \spa}{F^2}  \bra \{ P, \lam_i \} \big( \lam_j S 
\,u^\mu + u^\mu S \,\lam_j  \big)\ket
+\frac{\ssa \spa}{2F^2} \bra \big\{ S, \{ P, \lam_i \} \big\} 
\{ u^\mu, \lam_j \} \ket \nonumber \\
&&-\frac{c_d^2 \spa}{2F^4} \bra \{ S, \lam_i \} \big[[P, u^\mu],
 \lam_j \big]\ket -  \Big\{i\leftrightarrow j\Big\}\, , \\ \nonumber \\
\ks_{ij}&=&\delta_{ij}M_S^2 -\frac{\lambda^{\mathrm{SS}}_1}{2}\bra 
u^\mu u_\mu \{\lambda_i,\lambda_j\} \ket -\lambda^{\mathrm{SS}}_2 
\bra \lambda_i u_\mu \lambda_j u^\mu \ket - \frac{\lambda^{\mathrm{SS}}_3}{2} 
\bra \chi_+ \{\lambda_i , \lambda_j \} \ket \, , \\ \nonumber \\ 
\kp_{ij}&=&\delta_{ij}M_P^2 -\frac{\ppa}{2}\bra u^\mu u_\mu 
\{\lam_i,\lam_j\} \ket -\ppb \bra \lam_i u_\mu \lam_j u^\mu \ket 
- \frac{\ppc}{2} \bra \chi_+ \{\lam_i , \lam_j \} \ket \, ,  
\end{eqnarray}
\begin{eqnarray}
\sigma_{ij}&=&\frac{1}{16} \bra \chi_+  \{ \hat{\lambda}_i,\hat{\lambda}_j \}   \ket
-\frac{1}{16} \bra [ u_\mu,\hat{\lambda}_i][u^\mu,\hat{\lambda}_j]\ket 
\nonumber \\
&&-\frac{c_d}{4F^2} \bra \nabla^2 S \{ \lambda_i , \lambda_j \} \ket
+\frac{c_m}{8F^2} \bra \{ S,\hat{\lambda}_i\} \{ \chi_+, \hat{\lambda}_j \} 
\ket  
+\frac{c_d}{8F^2} \bra \{ S,u^\mu\} \big[ [u_\mu, \hat{\lambda}_i], 
\hat{\lambda}_j\big] \ket \nonumber \\
&&
-\frac{c_d \spa}{2F^2}\big(  \bra  \{ \nabla_\mu P, \hat{\lam}_i \} 
\{ u^\mu, \hat{\lam}_j \}  \ket +
 \bra  \{  P, \hat{\lam}_i \} \{\nabla_\mu u^\mu, \hat{\lam}_j \} 
  \ket \big) \nonumber \\
&&+\frac{i}{2F^2} \big( \frac{d_m}{4} + c_m \spa \big)\bra \{ P, 
\hat{\lam}_i \} \{ \chi_- , \hat{\lam}_j \} \ket
-\frac{1}{32F^2} \bra \big[ S,[ u^\mu,\lambda_i ]\big]  \big[S, 
[u_\mu,\lambda_j]\big] \ket  \nonumber \\
&&+\frac{1}{8F^2}  \bra[u^\mu ,\lambda_i ] 
 \big[ u_\mu , \tilde{\lambda}^{\mathrm{SS}}_1 \{ SS, \lambda_j 
 \}+2\tilde{\lambda}^{\mathrm{SS}}_2 S\lambda_j S \big) \big] \ket 
\nonumber \\
&&+\frac{1}{8F^2}  \bra[u^\mu ,\lambda_i ] 
\big[\lambda_j, \big( \lambda^{\mathrm{SS}}_1 \{ SS, u_\mu \}+2 
\lambda^{\mathrm{SS}}_2 S u_\mu S \big) \big] \Big) \ket 
\nonumber \\
&&-\frac{1}{8F^2}  \bra \{ \chi_+ ,\lambda_i \} \Big( \big(
 \tilde{\lambda}^{\mathrm{SS}}_1
-\lambda^{\mathrm{SS}}_3 \big)
 \{ SS, \lambda_j \}+2\tilde{\lambda}^{\mathrm{SS}}_2 S\lambda_j 
 S \Big) \ket  
-\frac{c_d^2}{2F^4} \bra \{ \nabla^\mu S, \lambda_i \} \{ \nabla_\mu S ,
 \lambda_j \} \ket\nonumber \\
&&-\frac{c_d^2}{4F^4} \bra \{S, \lambda_i  \} \{  \nabla^2 S , 
\lambda_j\} \ket
-\frac{\tilde{\lambda}^{\mathrm{SS}}_1}{4F^2} \bra \nabla^2 S^2 
\{ \lambda_i , \lambda_j \} \ket -\frac{\tilde{\lambda}^{\mathrm{SS}}_2}{F^2} 
\bra \lambda_i \nabla_\mu \big( \nabla^\mu S \lambda_j  S \big) \ket  
 \nonumber \\ 
&&+\frac{M_S^2 {(\spa)}^2}{2F^2} \bra \{ P, \lam_i \} \{ P, \lam_j \} 
\ket
-\frac{\tilde{\lam}^{\mathrm{PP}}_1}{8F^2} \bra \{ \chi_+ ,\lam_i \}\{ PP, 
\lam_j \} \ket
-\frac{\tilde{\lam}^{\mathrm{PP}}_2}{4F^2} \bra \{ \chi_+ ,\lam_i \}P\, 
\lam_j P  \ket \nonumber \\
&& -\frac{ \spa \spb}{ 2F^2} \bra \big\{ P, \{ P, \lam_i \} \big\} \{ \chi_+,
 \lam_j \} \ket \,  
+ \, \frac{\ppc}{8F^2} \bra \{ PP , \lam_i \} \{ \chi_+, \lam_j \} \ket 
\nonumber \\
&& -\frac{\ssc{(\spa)}^2}{F^2}\bra \{ P, \lam_i \} \{ P, \lam_j \} 
\chi_+ \ket \, 
-\,\frac{1}{32F^2} \bra \big[ P,[ u^\mu,\lam_i ]\big]  \big[P, [u_\mu,\lam_j]
\big] \ket \nonumber \\
&& +\frac{c_d^2 {(\spa)}^2}{4F^4} \bra \big[ [u^\mu,P],\lambda_i \big] \big[ 
[u_\mu,P], \lambda_j \big] \ket 
\,- \, \frac{{(\spa)}^2}{4F^2} \bra \big[ [u^\mu,\lambda_i], \{P, \lambda_j\} 
\big] \{P,u_\mu \} \ket \nonumber \\
&&-\frac{1}{8F^2} \bra \Big(\ppa \{ u_\mu, PP \}+2\ppb P \,u_\mu P \Big) 
\big[ [ \lam_i , u^\mu ], \lam_j \big] \ket \nonumber \\
&&- \, \frac{{(\spa)}^2}{F^2} \bra u^\mu \{ P, \lam_i \} \Big(  \ssa \{ P,
 \lam_j \} u_\mu +
 \ssb u_\mu \{ P, \lam_j \} \Big) \ket \nonumber \\
&&+\frac{\tilde{\lam}^{\mathrm{PP}}_1}{8F^2} \bra \big[ u_\mu , \{ PP, 
\lam_i \} \big] [u^\mu ,\lam_j ] \ket 
+\frac{\tilde{\lam}^{\mathrm{PP}}_2}{4F^2} \bra [ u_\mu , P\, \lam_i P 
\,] [u^\mu ,\lam_j ] \ket  \nonumber \\
&&-\frac{\tilde{\lam}^{\mathrm{PP}}_1}{4F^2} \bra \nabla^2 P^2 \{ \lam_i
 , \lam_j \} \ket 
-\frac{\tilde{\lam}^{\mathrm{PP}}_2}{F^2} \bra \lam_i \nabla_\mu \big(
 \nabla^\mu P \lam_j  P \big) \ket
 -\frac{{(\spa)}^2}{2F^2} \bra \{ \nabla_\mu P, \lam_i \} \{ \nabla^\mu P, 
 \lam_j \} \ket 
\nonumber \\
&& +\frac{i\spb}{8F^2} \bra \{ S,P \} \big\{ \lam_i, \{ \chi_-, \lam_j \} 
\big\} \ket 
 \,+ \, \frac{i\,\ssc \spa}{2F^2} \bra \big\{ S, \{ P, \lam_i \} \big\} \{ 
\chi_- ,\lam_j \} \ket \nonumber \\
&& -\frac{c_d^2 \spa}{2F^4} \bra \{ P, \lam_i \}  \big\{ u^\mu ,
 \{ \nabla_\mu S, 
\lam_j \} \big\} \ket \, - \, \frac{ \ssb \spa }{ F^2} \bra
 \nabla_\mu \Big( \{ P, \lam_i \} \big(
 \lam_j S u^\mu + u^\mu S \lam_j \big) \Big) \ket \nonumber \\
&&-\frac{ \ssa \spa}{2F^2} \bra \{ P, \lam_i \}  \nabla_\mu  \big\{ S , 
\{u^\mu , \lam_j \} \big\} \ket \nonumber \\
&&+\frac{\spa}{8F^2} \bra  [ u^\mu ,\lam_i] \Big(  \big[ \lam_j, 
\{ \nabla_\mu S, P \} \big] +
 2\big[ \nabla^\mu S, \{ P, \lam_j \} \big] -2 \big[ S , \{ \nabla^\mu P, 
 \lam_j \} \big] \Big) \ket  \nonumber \\
&&
+\frac{\spa}{2F^2} \bra \{ u^\mu , \lam_i \} \Big( \frac{c_d^2}{F^2} 
\big\{ P, \{ \nabla_\mu S, \lam_j \} \big\} - 
\ssa  \big\{ S, \{ \nabla_\mu P, \lam_j \} \big\} \Big) \ket 
+  \Big\{i\leftrightarrow j\Big\} \, , 
\end{eqnarray}
\begin{eqnarray}
\as_{ij}&=&-\frac{i\,c_m}{2\sqrt{2}F}\bra \chi_- \{\lambda_i,
\hat{\lambda}_j\} \ket
-\frac{1}{2\sqrt{2}F} \bra [\nabla^\mu S, \lambda_i ] [ u_\mu,
\hat{\lambda}_j ] \ket 
-\frac{1}{4\sqrt{2}F} \bra [S, \lambda_i ] [\nabla^\mu  u_\mu,
\hat{\lambda}_j ] \ket\nonumber \\
&&+\frac{c_d^2}{\sqrt{2}F^3} \bra \{u^\mu, \lambda_i \} \{\nabla_\mu S, 
\lambda_j \} \ket  
-\frac{i \, \lambda^{\mathrm{SS}}_3}{2\sqrt{2} F} \bra \{ S , \lambda_i 
\} \{ \chi_- , \hat{\lambda}_j \} \ket
-\frac{M_S^2 \spa}{\sqrt{2} F} \bra P \{ \lam_i , \hat{\lam}_j \} \ket   
\nonumber \\
&&
+ \frac{\spb}{2\sqrt{2}F} \bra \{ P, \lam_i \} \{ \chi_+ , \hat{\lam}_j 
 \} \ket 
+\frac{\spa \ssc }{\sqrt{2}F}\bra \{\chi_+,\lam_i\}\{P,\hat{\lam}_j \} 
\ket 
-\frac{\spa}{\sqrt{2}F}\bra \nabla^2 P \{ \lam_i, \hat{\lam}_j  \} 
\ket \nonumber \\
&&
 -\frac{\spa}{4\sqrt{2}F}\bra \{P, u^\mu \} \big[ \lam_i ,
 [u_\mu,\hat{\lam}_j ] \big] \ket
+\frac{\spa}{\sqrt{2}F} \bra \{P, \hat{\lam}_j \} \Big( \ssa 
\{u_\mu u^\mu, \lam_i\} +2\ssb u_\mu \lam_i u^\mu \Big) \ket 
\nonumber \\
&& +\frac{c_d}{\sqrt{2}F^3} \bra \{u^\mu, \lambda_i \} 
 \left( \tilde{\lambda}^{\mathrm{SS}}_1 \{  \nabla_\mu(SS), \lambda_j 
 \} +2 
\tilde{\lambda}^{\mathrm{SS}}_2 \nabla_\mu(S\lambda_j S) \right)  
\ket  
 \nonumber \\
&& +\frac{ \sqrt{2} \lambda^{\mathrm{SS}}_2 \, c_d}{F^3} \bra  \{ 
\nabla^\mu S, \lambda_j \} \Big(  S u_\mu \lambda_i  + \lambda_i u_\mu S 
\Big) \ket 
+\frac{ \lambda^{\mathrm{SS}}_1 \, c_d}{\sqrt{2} F^3} \bra \{ S, \lambda_i 
\} \big\{ u_\mu, \{ \nabla^\mu S, \lambda_j \} \big\} \ket 
 \nonumber \\
&& +\frac{c_d}{4\sqrt{2}F^3} \bra [ S, \lambda_i] \big[ u_\mu,\{ \nabla^\mu
 S,\lambda_j\}\big] \ket 
+\frac{i\, c_m }{2\sqrt{2} F^3} \bra \{ \chi_-,\lambda_i \}  \Big(\tilde{
\lambda}^{\mathrm{SS}}_1 \{ SS, \lambda_j \}+2\tilde{\lambda}^{\mathrm{SS}}_2
 S\lambda_j S \Big) \ket \nonumber \\
&& +\frac{i\, c_m}{2\sqrt{2}F^3} \bra \{ \chi_-, \lam_i\} \Big( 
\tilde{\lam}_1^{\mathrm{PP}} \{PP, \lam_j\}
+2\tilde{\lam}_2^{\mathrm{PP}}P\,\lam_j P \Big) \ket 
 \nonumber \\
&&+\frac{c_d}{\sqrt{2}F^3} \bra \{u^\mu, \lam_i \} 
\Big( \tilde{\lam}_1^{\mathrm{PP}} \{ \nabla_\mu (PP),\lam_j\} 
+2\tilde{\lam}_2^{\mathrm{PP}}\nabla_\mu (P\,\lam_j P) \Big) \ket \nonumber \\
&&
 +\frac{c_d\spa}{\sqrt{2}F^3} \bra  \{\nabla_\mu P,\lam_i\}\{\nabla^\mu S,
 \lam_j\} \ket \; , \\ \nonumber \\
\ap_{ij}&=& \frac{d_m}{2\sqrt{2}F} \bra \chi_+ \{\lam_i, \hat{\lam}_j \} 
\ket 
+\frac{\spb}{2\sqrt{2}F}\bra \{S,\lam_i\}\{\chi_+, \hat{\lam}_j\}\ket 
\nonumber \\
&& -\frac{\spa}{4\sqrt{2}F}\bra\{u_\mu,\lam_i\} \big[S,[u^\mu, \hat{\lam}_j]
\big]\ket
-\frac{1}{4\sqrt{2}F} \bra \Big( [P,\lam_i]  [\nabla^\mu u_\mu, \hat{\lam}_j 
] + 2 [\nabla^\mu P, \lam_i ]
[u_\mu, \hat{\lam}_j ] \Big) \ket \nonumber \\
&& +\frac{i\,\spa}{\sqrt{2}F} \spb\bra \{\chi_-,\lam_i\}\{P, 
\hat{\lam}_j\}\ket 
-\frac{i \, \ppc}{2\sqrt{2} F} \bra \{ P , \lam_i \} \{ \chi_- , 
\hat{\lam}_j \} \ket  \nonumber \\
&&
+\frac{{(\spa)}^2}{\sqrt{2}F} \bra \{u^\mu, \lam_i\}\{\nabla_\mu P, 
\hat{\lam}_j  \}\ket
+\frac{c_d\spa}{\sqrt{2}F^3}\bra \{\nabla_\mu S, \lam_i\} \{ \nabla^\mu S,
 \lam_j \} \ket \nonumber \\
&&-\frac{d_m}{2\sqrt{2} F^3} \bra \{ \chi_+,\lam_i \} \Big(  
\tilde{\lam}^{\mathrm{SS}}_1\{SS,\lam_j\} 
+2\tilde{\lam}^{\mathrm{SS}}_2 S\,\lam_j S 
+\tilde{\lam}^{\mathrm{PP}}_1\{ PP , \lam_j \} +2 
\tilde{\lam}^{\mathrm{PP}}_2 P \,\lam_j P \Big) \ket\,  \nonumber \\
&& -\frac{c_d {(\spa)}^2}{\sqrt{2}F^3} \bra \{u^\mu, \lam_i\} 
 \big\{P,\{\nabla_\mu S,\lam_j\} \big\} \ket
+\frac{c_d}{4\sqrt{2}F^3} \bra  [P,\lam_i] \big[ u_\mu,
\{\nabla^\mu S,\lam_j\}\big] \ket
\nonumber \\
&& +\frac{c_d }{\sqrt{2} F^3}\bra \{\nabla^\mu S,\lam_j\} \Big(
 \ppa \big\{ u_\mu ,\{P,\lam_i\} \big\}
+2\ppb \left( P \,u_\mu \lam_i +\lam_i u_\mu P \right) \Big) \ket  
 \, , \\ \nonumber \\
\asp_{ij}&=&\frac{i\,\spb}{2} \bra \chi_- \{ \lam_i,\lam_j\} \ket  
\, , 
%
\end{eqnarray}
\begin{eqnarray}
{{\bs}^{\mu}_{ij}}&=& -\frac{c_d}{\sqrt{2}F} \bra u^\mu \{ \lambda_i ,
\hat{\lambda}_j \} \ket
-\frac{1}{4\sqrt{2}F}\bra [S,\lambda_i][u^\mu,\hat{\lambda}_j]\ket 
-\frac{\lambda^{\mathrm{SS}}_1}{\sqrt{2}F} \bra \{ S, \lambda_i \} \{ 
u^\mu, \hat{\lambda}_j \} \ket 
 \nonumber \\
&&- \frac{\sqrt{2}\lambda^{\mathrm{SS}}_2}{F} \bra S \Big( u^\mu 
\lambda_i \hat{\lambda}_j + 
\hat{\lambda}_j \lambda_i u^\mu \Big) \ket 
-\frac{\spa}{\sqrt{2}F}\bra \nabla^\mu P \{\lam_i, \hat{\lam}_j \}
\ket \nonumber \\
&&+ \frac{c_d }{\sqrt{2} F^3} \bra \{ u^\mu , \lambda_i \} \Big( 
\tilde{\lambda}^{\mathrm{SS}}_1 \{ SS, \lambda_j \} +2
\tilde{\lambda}^{\mathrm{SS}}_2 S\lambda_j S 
+ \tilde{\lam}_1^{\mathrm{PP}} \{PP,\lam_j\} 
+2\tilde{\lam}_2^{\mathrm{PP}}P\lam_jP \Big) \ket \,  , \nonumber \\
 \\ \nonumber 
{{\bp}^{\mu}_{ij}}&=&-\frac{\spa}{\sqrt{2}F} \bra \nabla^\mu S
 \{\lam_i,\hat{\lam}_j \} \ket 
-\frac{1}{4\sqrt{2}F}\bra [P,\lam_i][u^\mu,\hat{\lam}_j]\ket  
-\frac{\ppa}{\sqrt{2}F} \bra \{ P, \lam_i \} \{ u^\mu, \hat{\lam}_j \} 
\ket \nonumber \\ 
&& - \frac{\sqrt{2}\ppb}{F} \bra P  \Big( u^\mu \lam_i \hat{\lam}_j +
\hat{\lam}_j \lam_i u^\mu\Big) \ket  
 +\frac{{(\spa)}^2}{\sqrt{2}F} \bra\{u^\mu,\lam_i\}\{P, \hat{\lam}_j\}
  \ket \, , \\ \nonumber \\
{{\bsp}^{\mu}_{ij}}&=& \frac{\spa}{2} \bra u^\mu \{\lam_i,\lam_j\} \ket \, ,
\end{eqnarray}
and the following definitions have been used,
\begin{equation}
\hat{\lambda}_i \,\equiv \, \lambda_i - \frac{c_d}{F^2} 
\{ \lambda_i , S \} \,, \qquad \qquad
\nabla_\mu \left( A \,\lambda_i\, B \right) \, \equiv \, 
\nabla_\mu A \, \lambda_i \, B \, + \, A \, \lambda_i \, \nabla_\mu B  \,,
\end{equation}
where $A$ and $B$ are any chiral tensor or resonance field.
\par
As commented in the text we can write Eq.~(\ref{eq:rchtapp}) as~:
\begin{equation}
\Delta {\cal L}_{\mathrm{R} \chi \mathrm{T}} \, = \, 
- \, \frac{1}{2} \, \eta \, \left( \, \Sigma_{\mu} \, \Sigma^{\mu} \, + \, 
\Lambda \, \right) \, \eta^{\top} \; , 
\end{equation}
where $\eta$ collects the fluctuation fields, 
$\eta=\left(\Delta_i,{\es}_j,{\ep}_k\right)$, $i,j,k = 0,...,8$, 
$\eta^{\top}$ is its transposed and $\Sigma_\mu$ and $\Lambda$ are defined as~:
\begin{eqnarray}
\left( \Sigma_\mu \right)_{ij} &=&\delta_{ij} \, \partial_\mu\, + \, 
\left( Y_{\mu} \right)_{ij} \, ,\\
& & \nonumber \\
\left( Y_{\mu} \right)_{ij} & = & \left( \begin{array}{ccc}
\gamma'_{\mu} & \frac{1}{2} {\bs_{\mu}}^{\top} &
\frac{1}{2} {\bp_{\mu}}^{\top}
\\\\ -\frac{1}{2} \bs_{\mu} & \gamma_{\mu} & \frac{1}{2} {\bsp_{\mu}}^{\top} 
\\\\
-\frac{1}{2} \bp_{\mu}  & -\frac{1}{2} \bsp_{\mu}& \gamma_{\mu}\end{array} 
\right)_{ij}\;\;\;,\label{eq:y}\\ \nonumber \\ 
& & \nonumber 
\end{eqnarray}
\begin{eqnarray}
\left( \Lambda \right)_{ij}  &=& \left( \begin{array}{ccc}
\sigma+ 
\frac{1}{4}{\bs_{\mu}}^{\top} \bsm  
& -\as^{\top}+\frac{1}{2}\tilde{d^{\mu}_-} {\bs_{\mu}}^{\top}
&-\ap^{\top}+\frac{1}{2}\tilde{d^{\mu}_-} {\bp_{\mu}}^{\top}
 \\
 \\
+ \frac{1}{4}{\bp_{\mu}}^{\top} \bpm 
& + 
\frac{1}{4}{\bp_{\mu}}^{\top} \bspm 
& - 
 \frac{1}{4}{\bs_{\mu}}^{\top} {\bspm}^{\top} 
\\ \\
\\
 -\as+\frac{1}{2} \bar{d^{\mu}_+} \bs_{\mu}
&\ks+ \frac{1}{4}\bsm {\bs_{\mu}}^{\top}  
& -\asp^{\top}+\frac{1}{2}\hat{d^{\mu}} {\bsp_{\mu}}^{\top}
\\
\\
 + \frac{1}{4}{\bsp_{\mu}}^{\top}
  \bpm 
& + \frac{1}{4}{\bsp_{\mu}}^{\top} 
\bspm 
&  + 
\frac{1}{4}\bsm{\bp_{\mu}}^{\top} 
\\ \\ \\
-\ap+\frac{1}{2} \bar{d^{\mu}_+} \bp_{\mu} 
&-\asp+\frac{1}{2} \hat{d^{\mu}} \bsp_{\mu} 
& \kp+ \frac{1}{4}\bpm {\bp_{\mu}}^{\top}  \\ \\
 - \frac{1}{4}\bspm \bs_{\mu} 
& + \frac{1}{4}\bpm 
{\bs_{\mu}}^{\top} 
& + \frac{1}{4}\bspm
  {\bsp_{\mu}}^{\top} 
 \end{array} \right)_{ij} \; .\nonumber \\ 
\end{eqnarray}
Here some new expressions have been defined~:
\begin{eqnarray} \label{eq:gamos}
\gamma^{\mu} &=&{\gamma^\mu}\big|_{\chi} \, , \nonumber \\
\gamma'^{\mu} &=& {\gamma^\mu}\big|_{\chi} + {\gamma^\mu}\big|_{\mathrm{R}} 
\, ,\nonumber \\ 
\hat{d^{\mu}} \, X &=& \partial^{\mu} \, X \, + \, \left[ \, \gamma^{\mu}  \, ,
X \, \right] \, , \nonumber  \\
\tilde{d^{\mu}_{\pm}} X &=& \hat{d^{\mu}} \,  X \, \pm \, 
\left(\gamma'^{\mu}-\gamma^{\mu}\right)\,X \,,\nonumber  \\
\bar{d^{\mu}_{\pm}} X &=& \hat{d^{\mu}} \, X \, \pm \, X\,
\left(\gamma'^{\mu}-\gamma^\mu \right) \, .
\end{eqnarray}

\appendix
\newcounter{ovidio}
\renewcommand{\thesection}{\Alph{ovidio}}
\renewcommand{\theequation}{\Alph{ovidio}.\arabic{equation}}
\renewcommand{\thetable}{\Alph{ovidio}.\arabic{table}}
\setcounter{ovidio}{5}
\setcounter{equation}{0}
\setcounter{table}{0}

\section{Results~: Operators and coefficients of the $\beta$-function}
\label{app:cinco}

In this Appendix we collect part of our results. In Table~\ref{tab:e1}
we have written the operators that involve only Goldstone fields
and/or external
currents. The coefficients of the $\beta$-function are generated by 
full ${\cal L}_{\mathrm{R} \chi \mathrm{T}}$, i.e. including both
scalar and pseudoscalar resonances. They can be compared 
with the study of one-loop renormalization
of $U(3)_L \otimes U(3)_R$ $\chi$PT up to ${\cal O}(p^4)$ in 
Ref.~\cite{Herrera-Siklody:1996pm}. The operators in this table constitute
a minimal basis. If one wishes to disentangle the pseudoscalar 
contributions one has to cancel the corresponding
resonance couplings and substitute
$N/16 \rightarrow N/24$ in $\gamma_4$, $0 \rightarrow N/48$ in 
$\gamma_7$, $N/4 \rightarrow N/6$ in $\gamma_{13}$, $-N/4 \rightarrow
-N/6$ in $\gamma_{14}$ and $-N/8 \rightarrow -N/12$ in $\gamma_{15}$.
\par
In Tables~\ref{tab:e2}, \ref{tab:e3} we collect the operators involving
one and two scalar resonances, respectively. The $\beta$-function 
coefficients are those obtained by considering 
Goldstone and scalar resonances only. 
The full result when pseudoscalar resonances are also introduced
in the theory can be looked up at
\texttt{http://ific.uv.es/quiral/rt1loop.html} 
or upon request from the authors. We would like to emphasize that
the basis of operators has been simplified by the use of the EOM in 
Appendix~\ref{app:tres} but possible $U(3)$ algebraic relations have
not been employed. Accordingly our basis is not necessarily minimal.
In the tables $N$ is the number of flavours and $u\cdot u \equiv u_\mu u^\mu$.
{\small
\begin{center}
\tabletail{\hline}
\topcaption{\label{tab:e1} Operators involving only Goldstone bosons and
 external currents and
their $\beta$-function coefficients at one loop, when both scalar and
pseudoscalar resonances are included. }
\begin{supertabular}{|>{$}r<{$} | >{$}c<{$} | >{$}p{12cm}<{$} |}
\hline & & \\
i & \cO_i &\qquad \qquad \qquad \qquad \qquad \qquad \quad \gamma_i  \\ & & \\ \hline \hline 

1&       \bra  u \cdot u \ket &   - 2    N        {\ppa}    {M_P^2} + 1/2    N        {M_P^2}    (\spa)^2 - 2    N        {\ssa}    {M_S^2}
          + 1/2    N        {M_S^2}    (\spa)^2 + N  F^{-2}  c_d^2    {M_S^2} \\ \hline

2&       \bra  {\chi_+} \ket &   - 2    N        {\ppc}    {M_P^2} - 2    N        {\ssc}    {M_S^2} \\ \hline

3&      - \bra  u_\mu \ket^2 &    2        {\ppb}    {M_P^2} - 1/2        {M_P^2}    (\spa)^2 + 2        {\ssb}    {M_S^2}
          - 1/2        {M_S^2}    (\spa)^2 - c_d^2  F^{-2}  {M_S^2} \\ \hline

4&       \bra  u_\mu   u_\nu u^\mu u^\nu   \ket &  1/6    N    F^{-4}    c_d^4 - 1/12    N        (\spa)^2 + 1/24    N
                 (\spa)^4 + 1/16    N     + 1/6    N  F^{-2}  c_d^2    (\spa)^2 - 1/6    N  F^{-2}  c_d^2 \\ \hline

5&       \bra  u \cdot u\ket^2 &  1/2    F^{-4}    c_d^4 - 1/2        {\ppa}    (\spa)^2 +     (\ppa)^2 - 1/2        {\ssa}
             (\spa)^2 +     (\ssa)^2 + 1/8        (\spa)^4 + 1/16     - F^{-2}c_d^2    {\ssa} + 1/2  F^{-2}  c_d^2
         (\spa)^2 - 1/4  F^{-2}  c_d^2 \\
\hline
\end{supertabular}
\end{center}}
\vspace*{0.5cm}
%
{\small \begin{center}
\tabletail{\hline}	 
\topcaption{\label{tab:e2} Operators with one scalar 
resonance and their $\beta$-function 
coefficients when only scalar resonances are included.}
\begin{supertabular}{|>{$}r<{$} | >{$}c<{$} | >{$}p{11.5cm}<{$} |}
\hline & & \\
i & \cO_i^{S} & \qquad \qquad \qquad \qquad \qquad \qquad \quad \gamma_i^{S} \\ & & \\ \hline \hline 
1 &

       \bra S \, u \cdot u  \ket &   - 3 \, N \, F^{-4} \, c_d^3 \, {M_S^2} + 2 \, N \, F^{-2} \, c_d \,
       {\ssa} \, {M_S^2} + 4 \, N \,  F^{-2} \,c_d \, {\ssb} \, {M_S^2} \\ \hline 2 &

       \bra S \ket \bra u \cdot u  \ket &   - 3 \, F^{-4} \, c_d^3 \, {M_S^2} + 2 \,  F^{-2} \,c_d \, {\ssa}
       \, {M_S^2} + F^{-2} \, c_d \, {M_S^2} \\ \hline 3 &

      \bra u^\mu  \ket \bra u_\mu  \, S \ket &   - 2 \, F^{-4} \, c_d^3 \, {M_S^2} + 4 \, F^{-2} \, c_d \, {\ssa} \, {M_S^2} + 4 \,  F^{-2} \,c_d
          \, {\ssb} \, {M_S^2} -  F^{-2} \,c_d \, {M_S^2} \\ \hline 4 &

       \bra S \, \chi_+  \ket &   - 2 \, N \, F^{-4} \, c_d^2 \, c_m \, {M_S^2} + 1/2 \, N \,  F^{-2} \,c_d \,
       {M_S^2} \\ \hline 5 &

       \bra S \ket \bra \chi_+  \ket &   - 2 \, F^{-4} \, c_d^2 \, c_m \, {M_S^2} + 1/2 \, F^{-2} \, c_d \,
       {M_S^2} \\
\hline 
\end{supertabular}
\end{center}}
\vspace*{1cm}
%
{\small \begin{center}
\tabletail{\hline}
\tablecaption{\label{tab:e3} Operators with two scalar resonances
 and their $\beta$-function 
coefficients when only scalar resonances are included.}
\begin{supertabular}{|>{$}r<{$} | >{$}c<{$} | >{$}p{11.2cm}<{$} |} \hline & & \\
i & \cO_i^{SS} &\qquad \qquad \qquad \qquad \qquad \qquad \quad \gamma_i^{SS} \\ & & \\ \hline \hline 
1 &

       \bra SS \ket &  N \, F^{-4} \, c_d^2 \, {M_S^4} \\ \hline 2 &

       \bra S \ket^2 &  F^{-4} \, c_d^2 \, {M_S^4} \\ \hline 3 &

       \bra SS \, u \cdot u  \ket &   - 4 \, N \, F^{-4} \, c_d^2 \, {\ssa} \, {M_S^2} - 4 \, N \, F^{-4} \, c_d^2 \, {\ssb} \, {M_S^2} - N \, F^{-4}
          \, c_d^2 \, {M_S^2} + 1/2 \, N \, F^{-2} \, {\ssa} \, {M_S^2} + N \,  F^{-2} \,(\ssa)^2 \, {M_S^2} +
	  4 \, N \, F^{-2} \, (\ssb)^2 \, {M_S^2} + 1/16 \, N \, F^{-2} \, {M_S^2} \\ \hline 4 &

       \bra u_\nu S u^\nu  \, S \ket &   - 2 \, N \, F^{-4} \, c_d^2 \, {\ssa} \, {M_S^2} - 4 \, N \, F^{-4} \, c_d^2 \, {\ssb} \, {M_S^2} + 3/2 \, N \, 
         F^{-4} \, c_d^2 \, {M_S^2} + 4 \, N \, F^{-2} \, {\ssa} \, {\ssb} \, {M_S^2} - N \,  F^{-2} \,{\ssb}
	 \, {M_S^2} \\ \hline 5 &

       \bra u^\mu  \ket \bra u_\mu  \, SS \ket &   - 4 \, F^{-4} \, c_d^2 \, {\ssa} \, {M_S^2} - 4 \, F^{-4} \, c_d^2 \, {\ssb}
          \, {M_S^2} + 2 \, F^{-4} \, c_d^2 \, {M_S^2} + 4 \, F^{-2} \, {\ssa} \, {\ssb} \, {M_S^2} - 2 \, F^{-2} \, {\ssa} \, {M_S^2} + 2 \, F^{-2} \, (\ssa)^2 \, {M_S^2} + F^{-2} \, {\ssb} \, {M_S^2}
          - 1/8 \, F^{-2} \, {M_S^2} \\
\hline 

\end{supertabular}
\end{center}}


\end{document}